\documentclass[a4paper,12pt]{article}
\usepackage{jcappub} 
\usepackage{cite}
\usepackage{graphicx}
\usepackage{enumerate}
\usepackage{cancel}
\usepackage{color}
\usepackage{graphicx}
\usepackage{amsmath}
\usepackage{amsfonts}
\usepackage{amssymb}
\usepackage{rotating}
\usepackage{booktabs}
\usepackage{xcolor}
\usepackage{soul}

\def\comment#1{}

\title{\boldmath 
Spontaneous Peccei-Quinn symmetry breaking renders sterile neutrino, axion and $\chi$boson to be candidates for dark matter particles
}

\author{She-Sheng Xue}

\affiliation{ICRANet Piazzale della Repubblica, 10 -65122, Pescara, Italy, \\ Physics Department, University of Rome La Sapienza, \\ P.le Aldo Moro 5, I–00185 Rome, Italy\\
INFN, Sezione di Perugia, Via A. Pascoli, I-06123, Perugia, Italy
}

\emailAdd{xue@icra.it and shesheng.xue@gmail.com} 

\abstract{We study the Peccei-Quinn (PQ) symmetry of the sterile right-handed neutrino sector and the gauge symmetries of the Standard Model. Due to four-fermion interactions, spontaneous breaking of these symmetries at the electroweak scale generates top-quark Dirac mass and sterile-neutrino Majorana mass. The top quark channel yields massive Higgs, $W^\pm$ and $Z^0$ bosons. The sterile neutrino channel yields the heaviest sterile neutrino Majorana mass, sterile Nambu-Goldstone
axion (or majoron) and massive scalar $\chi$boson. Four-fermion operators effectively induce their tiny couplings to SM particles. We show that a sterile QCD axion is the PQ solution to the strong CP problem. The lightest and heaviest sterile neutrinos ($m_N^e\sim 10^2$ keV and $m_N^\tau\sim 10^2$ GeV), a sterile QCD axion 
($m_a< 10^{-8}$ eV, $g_{a\gamma}< 10^{-13} {\rm GeV}^{-1}$) and 
a Higgs-like $\chi$boson ($m_\chi\sim 10^2$ GeV) can be dark matter particle candidates, for the constraints of their tiny couplings and long lifetimes inferred from the
$W$-boson decay width, Xenon1T and precision fine-structure-constant 
experiments. The axion and $\chi$boson couplings to SM particles are below the values reached by current laboratory experiments and astrophysical observations for directly or indirectly detecting dark matter particles.}

\begin{document}
\maketitle
\flushbottom

%
\section
{\bf Introduction}\label{int}

Over eight decades, evidence has been 
built up by astrophysical and cosmological observations 
that cannot be explained unless the dark matter is present in addition to normal matter made up of fundamental particles in the Standard Model (SM) of particle physics \cite{Spergel2015,Tanabashi2018}. 
Dark matter accounts for approximately $85\%$ 
of the total matter in the universe. These facts constitute physically compelling reasons for new fundamental particles and interactions beyond SM. There is a broad range of dark matter candidates since evidence has only been observed so far through gravitational effects over large length scales,
from the size of the largest superclusters of galaxies down to the smallest observable dwarf galaxies.

The weakly-interacting massive particle (WIMP), axion-like particles (APLs) 
and sterile right-handed neutrinos are theoretically well-motivated dark matter particle candidates. WIMP candidates are of typical $10^2$ GeV masses and electroweak interactions. They are produced thermally in the early Universe and give the correct abundance of dark matter today (WIMP miracle) \cite{Donato2009,Arcadi2018,Slatyer2009,Roszkowski2018,Conrad2015,Steigman2012}. Supersymmetric extensions of the SM predict a new particle with these properties \cite{Jungman1996}. The Peccei-Quinn (PQ) axion \cite{Peccei1977,Peccei1977a,Weinberg1978,Wilczek1978} offers a compelling solution to the strong CP problem of quantum chromodynamics \cite{Preskill1983,Davidson1982,Abbott1983,Dine1983,DeMille2017,Irastorza2018}. The axion and other light ALPs 
emerge naturally from theoretical models of physics at high energies, including string theory, grand unified theories, and models with extra dimensions \cite{Irastorza2018,Svrcek2006}. The extra neutrinos are added in the $\nu$MSM, left-right symmetry and other extensions of the SM model \cite{PhysRevD.11.2558,PhysRevD.12.1502,Kusenko:2009up,Nemevsek2012,Drewes2013,Jana:2019mez,Eichten1986}. The right-handed neutrinos $\nu_R$ 
and four-fermion interactions have to be 
present \cite{Xue1996,Xue2000,Xue:1999xa,Xue:2001he}, due to chiral gauge 
symmetries of SM cannot be simply consistent with a cutoff field 
theory \cite{Nielsen:1981xu,1981PhLB..105..219N}. 

There is a large number of ultra-sensitive experiments searching for WIMP particles \cite{Tanabashi2018,Rajendran2017,Liu2017}.  
Astrophysical observations and laboratory experiments have produced a number of stringent limits on ALPs \cite{Gramolin2020,Ouellet2018,Anastassopoulos:2017ftl,Matsuura2017,Kohri2017,Duffy2009a,Arvanitaki2019,Giagu2019}. 
The recent Xenon1T \cite{Aprile2020a} experiment possibly sheds new light on sterile neutrinos as dark matter particles \cite{Shakeri2020}. However, so far there has been no unambiguous direct or indirect detection of these 
dark matter particles.

There are several ways to probe into dark matter particle candidates, 
we adopt an effective field theory, analogously to 
the approach \cite{Beltran2008,Beltran2010}.  
In the previous articles \cite{Xue2015,Xue:2016dpl}, we preliminarily study the spontaneous breaking of global $U(1)$ chiral symmetry in the sterile right-handed neutrino sector, which leads to three possible DM (dark matter) particle candidates: massive sterile neutrinos, pseudoscalar and scalar bosons. 
In the recent article \cite{Shakeri2020}, the right-handed neutrinos masses and coupling ${\mathcal G}_R$ to SM gauge bosons have been inferred	 
by the Xenon1T experiment \cite{Aprile2020a} and constrained by astrophysical observations \cite{Raffelt:1994ry,Arceo-Diaz:2015pva}. From these results, we can estimate three dark matter particle candidates' masses and couplings to SM particles. 
We find that sterile neutrinos Majorana masses are generated by spontaneous $U(1)$ symmetry breaking, accompany by a pseudoscalar Goldstone boson of PQ type axion $a$ (or Majoron) and massive scalar $\chi$boson of mass ${\mathcal O}(10^2)$ GeV. They can be potential DM particle candidates, for their very long lifetimes and tiny couplings to SM particles.

The article is arranged as follow. In Sec.~\ref{quadrilinear}, we briefly introduce the physical compelling reasons for right-handed neutrinos and their four-fermion interactions in an effective theory at an ultraviolet 
(UV) cutoff. Similarly to the analysis of top-quark condensate and spontaneous SM electroweak gauge symmetries breaking in Sec.~\ref{SSBS}, we study the spontaneous breaking of sterile neutrino PQ symmetry, the sterile neutrino condensate and Majorana masses in Secs.~\ref{sterile} and \ref{warmdm}. 
We present detailed discussions for the sterile QCD axion at the electroweak scale in Sec.~\ref{axion}, and the Higgs-analog sterile $\chi$boson in Sec.~\ref{Xboson}. Summary and remarks are in the last section Sec.~\ref{con}.

\section{Sterile neutrinos and four-fermion interactions}\label{quadrilinear}

A well-defined quantum field theory for the SM Lagrangian requires a natural regularization (UV cutoff $\Lambda_{\rm cut}$) fully preserving the SM 
chiral-gauge symmetry. The UV cutoff $\Lambda_{\rm cut}$ could be of the Planck scale or grand unified theory (GUT) scale. Quantum gravity or other new physics naturally provides such regularization.
However, the no-go theorem \cite{Nielsen:1981xu,1981PhLB..105..219N} shows the presence of right-handed neutrinos and absence of consistent ways to regularize the SM bilinear fermion Lagrangian to exactly preserve the SM chiral-gauge symmetries. This implies SM fermions' and right-handed neutrinos' four-fermion operators at the UV cutoff. As a theoretical model, we adopt the four-fermion operators of the torsion-free Einstein-Cartan Lagrangian with SM fermion content and three right-handed neutrinos \cite{Xue2015,Xue:2016dpl,Xue:2016txt},
\begin{eqnarray}
{\mathcal L}
&\supset &-G\sum_{f}\left(\, \bar\psi^{f}_{_L}\psi^{f}_{_R}\bar\psi^{f}_{_R} \psi^{f}_{_L}
+\, \bar\nu^{fc}_{_R}\psi^{f}_{_R}\bar\psi^{f}_{_R} \nu^{fc}_{_R}\right)+{\rm h.c.},
\label{art1}
\end{eqnarray}
where the two component Weyl fermions $\psi^{f}_{_L}$ and $\psi^{f}_{_R}$  
respectively are the eigenstates of the SM gauge symmetries $SU_C(3)\times SU_L(2)\times U_Y(1)$. For the sake of compact notations, $\psi^{f}_{_R}$ is also used to represent 
three right-handed sterile neutrinos $\nu^f_R$, which are SM gauged singlets. 
All fermions are massless, they are four-component Dirac fermions 
$\psi^f=(\psi_L^f+\psi_R^f)$, two-component left-handed Weyl neutrinos $\nu^f_L$ and four-component sterile Majorana neutrinos $\nu_M^f=(\nu_R^{fc}+\nu_R^f)$, where $\nu_R^{f c}=i\gamma_2(\nu_R^{f})^*$. 
In Eq.~(\ref{art1}), $f=1,2,3$ are 
fermion-family indexes summed over respectively for three 
lepton families (charge $q=0,-1$) and three quark families ($q=2/3,-1/3$). 
Eq.~(\ref{art1}) preserves not only the SM gauge symmetries and global fermion-family symmetries but also the global $U(1)$ symmetries for conservations of fermion numbers. We adopt the effective four-fermion operators (\ref{art1}) and coupling $G\propto {\mathcal O}(\Lambda_{\rm cut}^{-2})$ in the context of a well-defined quantum field theory at the high-energy scale $\Lambda_{\rm cut}$. Here, we suppose that the four-fermion coupling $G$ is the unique coupling, implying effective operators 
(\ref{art1}) at the UV cutoff are the same for all SM fermions. The assumptions are (i) such Einstein-Cartan type operators should be attributed to the nature of quantum gravity at the UV cutoff; (ii) all SM fermions are massless, and their eigenstates of mass and gauge interactions are the same at the UV cutoff. However, 
this is a preliminary approximation in the effective Lagrangian. 
The SM family mixing must occur in the ground state when SM fermion mass and gauge eigenstates are different. Thus the effective four-fermion coupling $G$ should be different for different SM fermion species, due to family mixing angles that are treated as parameters. We will duly address this issue later.

Among operators in the Lagrangian (\ref{art1}), 
we explicitly show the operators relevant to the issues of this article.
In the first term of Eq.~(\ref{art1}), the top-quark channel 
interaction is given by Bardeen, Hill and Lindner (BHL) \cite{Bardeen1990,Cvetic1999} 
\begin{eqnarray}
G(\bar\psi^{ia}_Lt_{Ra})(\bar t^b_{R}\psi_{Lib}),
\label{bhl}
\end{eqnarray}
where $a$ and $b$ are the color indexes 
of the top and bottom quarks, the quark $SU_L(2)$ doublet 
$\psi^{ia}_L=(t^{a}_L,b^{a}_L)$ 
and singlet $t^{a}_R$ are the eigenstates 
of SM electroweak interaction. 
Coming from the second term ($\psi^{f\,}_{R}=\nu^{\ell\,}_{R}$) 
in Eq.~(\ref{art1}), the sterile-neutrinos channel $\nu^\ell_R$ ($\ell=e,\mu,\tau$) is,
\begin{eqnarray}
G(\bar\nu^{\ell\, c}_R\nu^{\ell\,}_{R})(\bar \nu^{\ell}_{R}\nu^{\ell c}_{R}),
\label{bhlxl}
\end{eqnarray} 
which preserves the global chiral symmetry 
$U_{\rm lepton}(1)$ for the $\nu^{\ell}_{R}$ lepton-number 
conservation, although $(\bar \nu^{\ell}_{R}\nu^{\ell c}_{R})$ 
violates the lepton number of family ``$\ell$'' by two units. 

\section
{Spontaneous breaking of SM gauge symmetries}\label{SSBS}

Apart from what is possible new physics at the UV scale $\Lambda_{\rm cut}$ explaining the 
origin of these effective four-fermion operators (\ref{art1}), 
it is essential and necessary to study: (i) which dynamics of these operators 
undergo in terms of their couplings as functions of running energy 
scale $\mu$; (ii) associating to these dynamics where the infrared (IR) 
or ultraviolet (UV) stable fixed point of physical couplings locates; 
(iii) in the domains (scaling regions) of these stable fixed points, 
which physically relevant operators that 
become effectively dimensional-4 renormalizable 
operators following RG equations (scaling laws), 
while other irrelevant operators are suppressed by the cutoff at least 
$\mathcal O(\Lambda_{\rm cut}^{-1})$.    

\begin{figure}[t]
\centering
\includegraphics[width=1.5in]{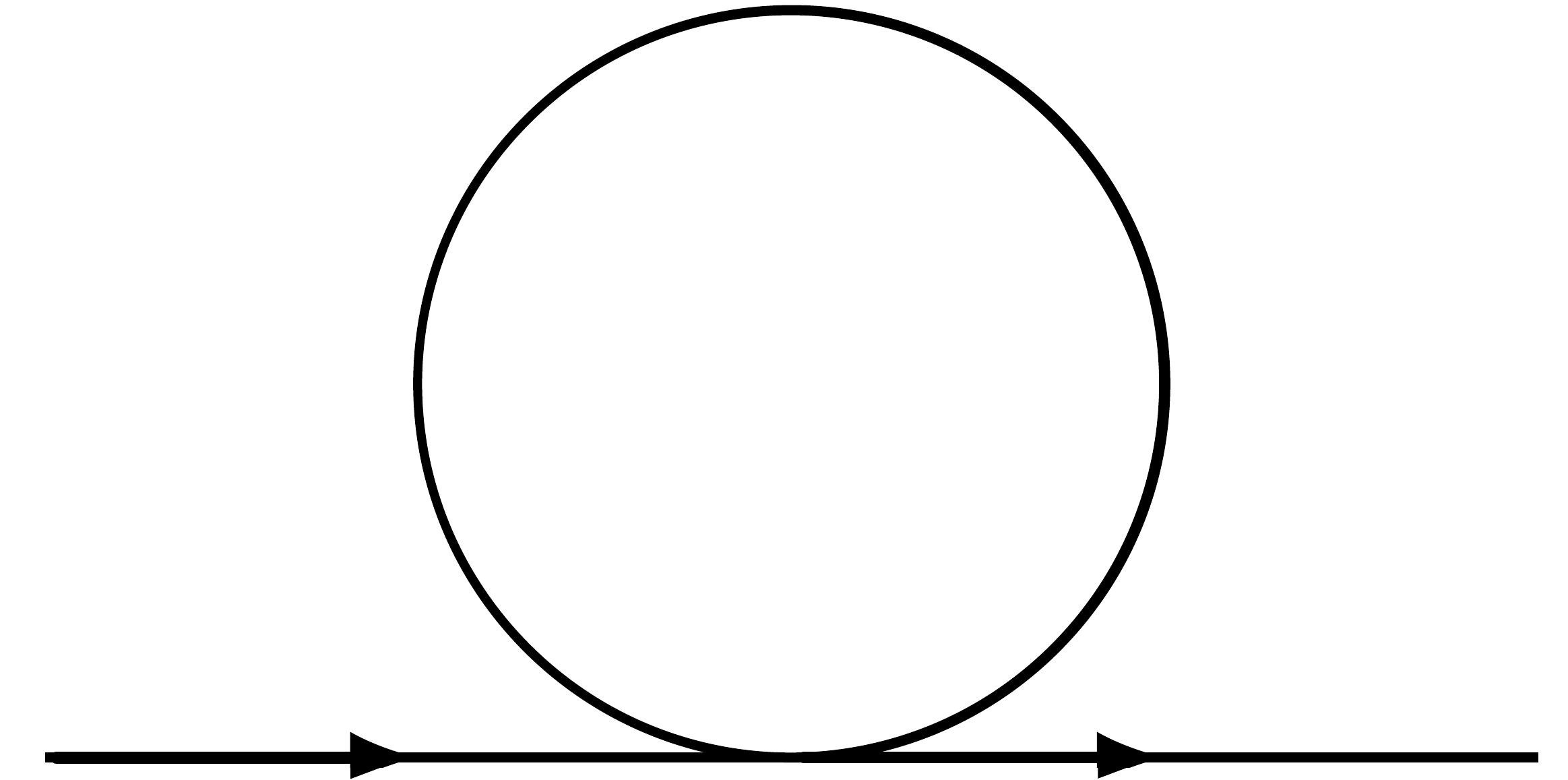}\hskip2.0cm\includegraphics[width=1.5in]{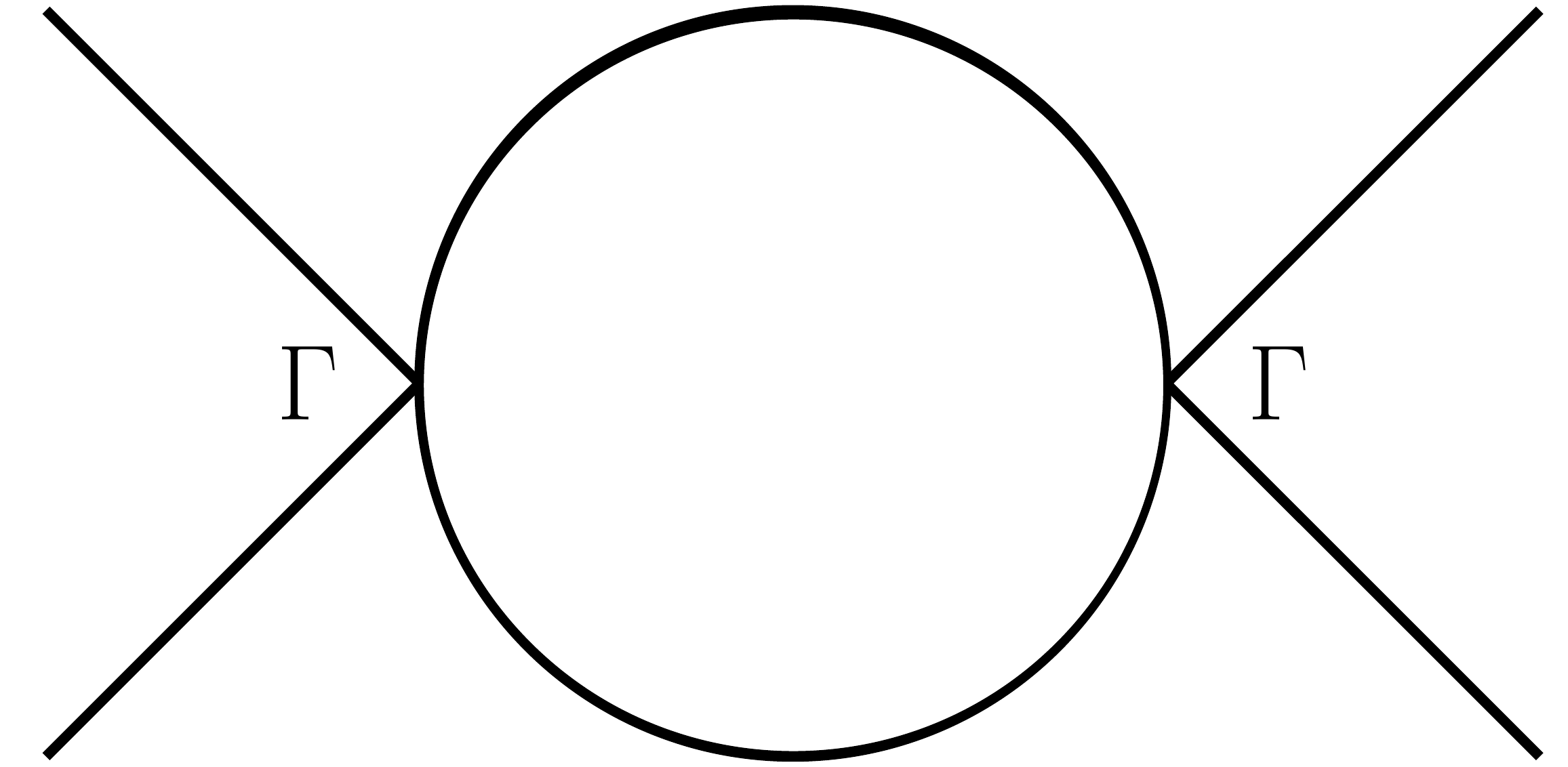}
    \caption{Left: The tadpole diagram stands for the gap equation in SSB. Right: a bubble diagram, where $\Gamma =1, \gamma_5$ for scalar or 
pseudoscalar coupling vertexes. The solid lines and circle indicate sterile neutrino (or top quark) propagators and loop. The four sterile neutrinos 
(top quarks) interacting vertex is associated with the coupling strength $G/2$. The indications are the same in the following figures, unless otherwise stated.}
    \label{tadpole}
\end{figure}

\begin{figure}[t]
\centering
\includegraphics[width=6.5in]{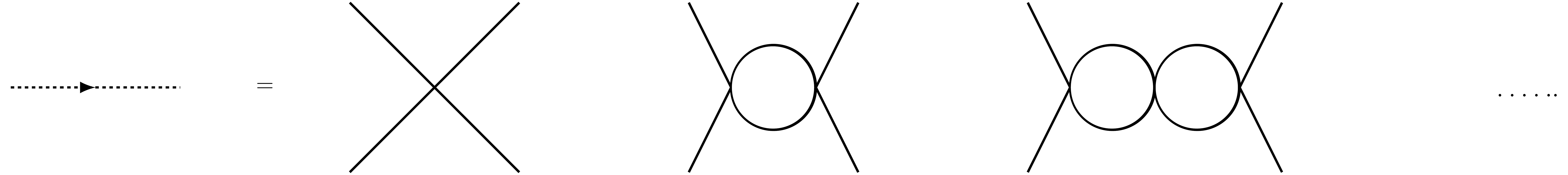}
\put(-60,22){+}\put(-190,22){+}\put(-290,22){+}
\caption{The diagram of summing bubbles represents composite scalar boson or pseudo scalar boson (dashed line).}
    \label{bubblesum}
\end{figure}

\subsection{Top-quark 
channel 
and low-energy effective theory}\label{top}

In the domain of the IR-stable fixed point $G_c$, using the 
approach of large-$N$ expansion,
it was shown \cite{Bardeen1990} that the 
operator (\ref{bhl}) undergoes the spontaneous symmetry 
breaking (SSB) dynamics, leading to the generation of top-quark mass  
\begin{eqnarray}
m_t&=&-(G/N_c)\sum_a\langle \bar t^a_L t_{aR}\rangle=2 G\frac{i}{(2\pi)^4}\int^{\Lambda_{\rm cut}} d^4p \frac{m_t}{(p^2-m_t^2)}.
\label{tqmassd}
\end{eqnarray}
The mass gap-equation, see 
tadpole diagram of Fig.~\ref{tadpole}, removes the $\Lambda^2_{\rm cut}$-divergence and obtain $m_t \ll \Lambda_{\rm cut}$ by fine tuning 
$G$ value.  
It appears the composite Higgs scalar 
$\langle\bar tt(x)\rangle$ and  
Nambu-Goldstone bosons, e.g., $\langle\bar t(x)\gamma_5t(x)\rangle$, see Fig.~\ref{bubblesum}. The latter 
becomes the longitudinal modes of the massive $Z^0$ and $W^\pm$ gauge bosons.
The effective SM Lagrangian with the {\it bilinear} top-quark mass term and 
Yukawa coupling to the composite Higgs boson $H$ at the low-energy 
scale $\mu$ is given by \cite{Bardeen1990,Cvetic1999}
\begin{eqnarray}
L &=& L_{\rm kinetic} + g_{t0}(\bar \Psi_L t_RH+ {\rm h.c.})
\nonumber\\ 
&+& Z_H|D_\mu H|^2-m_{_H}^2H^\dagger H
-\frac{\lambda_0}{2}(H^\dagger H)^2,
\label{eff}
\end{eqnarray}
where the bare Yukawa coupling
$g_{t0}$, static Higgs mass $m_0(m_{_H})\approx \Lambda_{\rm cut}$ and quartic coupling 
$\lambda_0$ at the UV cutoff scale $\Lambda_{\rm cut}$, and the finite coupling $G$
is given by $G=g^2_{t0}/m^2_0$, see Eqs.~(3-1-3.3) of Ref.\cite{Bardeen1990}.  
All renormalized quantities received fermion-loop contributions are 
defined with respect to the low-energy scale $\mu$. 
The conventional renormalization $Z_\psi=1$ for fundamental 
fermions and the unconventional wave-function renormalization (form factor)
$\tilde Z_H$ for the composite Higgs boson are 
adopted
\begin{equation}
\tilde Z_{H}(\mu)=\frac{1}{\bar g^2_t(\mu)},\, \bar g_t(\mu)=\frac{Z_{HY}}{Z_H^{1/2}}g_{t0}; \quad \tilde \lambda(\mu)=\frac{\bar\lambda(\mu)}{\bar g^4_t(\mu)},\,\bar\lambda(\mu)=\frac{Z_{4H}}{Z_H^2}\lambda_0,
\label{boun0}
\end{equation}
where $Z_{HY}$ and $Z_{4H}$ are proper renormalization constants of 
the Yukawa coupling and quartic coupling in Eq.~(\ref{eff}). 
In the IR-domain 
where the SM particle physics is realized at the electroweak energy scale $v=2^{-1/4}G_F^{-1/2}\approx246$ GeV, 
the full one-loop renormalization group (RG) equations for running couplings $\bar g_t(\mu^2)$ and $\bar \lambda(\mu^2)$
read
\begin{eqnarray}
16\pi^2\frac{d\bar g_t}{dt} &=&\left(\frac{9}{2}\bar g_t^2-8 \bar g^2_3 - \frac{9}{4}\bar g^2_2 -\frac{17}{12}\bar g^2_1 \right)\bar g_t,
\label{reg1}\\
16\pi^2\frac{d\bar \lambda}{dt} &=&12\left[\bar\lambda^2+(\bar g_t^2-A)\bar\lambda + B -\bar g^4_t \right],\quad t=\ln\mu \label{reg2}
\end{eqnarray}
where one can find $A$, $B$ and RG equations for 
SM $SU_c(3)\times SU_L(2)\times U_Y(1)$ running  gauge couplings $\bar g^2_{1,2,3}$ in Eqs.~(4.7), (4.8) of 
Ref.~\cite{Bardeen1990}. The SSB-generated top-quark mass $m_t(\mu)=\bar g_t^2(\mu)v/\sqrt{2}$. 
The composite Higgs-boson is described by its pole-mass 
$m^2_H(\mu)=2\bar \lambda(\mu) v^2$, form-factor $\tilde Z_H(\mu)=1/\bar g_t^2(\mu)$ and effective quartic coupling $\tilde\lambda(\mu)$, provided that 
$\tilde Z_H(\mu)>0$ and $\tilde\lambda(\mu)>0$ are obeyed.
As a result, the heaviest top quark mass is generated 
by the spontaneous breaking of SM gauge symmetries in the top sector (\ref{bhl})  
with a $t\bar t$ bound state as a candidate for the SM Higgs particle, 
and three Goldstone bosons becoming the longitudinal modes of massive gauge bosons 
$W^\pm$ and $Z^0$. 
This scenario provides a low energy effective theory for the SM.

In Refs.~\cite{Preparata1996,Xue2013c}, we explain that 
among four-fermion operators (\ref{art1}), 
why only the top-quark sector (\ref{bhl}) undergoes 
the condensation and the top quark mass is generated by spontaneously symmetry breaking. 
The reason is that the top-quark condensation gives the least numbers of Goldstone bosons, 
and this is the energetically favourable ground state of four-fermion interactions (\ref{art1}).
Other fermion Dirac masses are generated by induced explicitly symmetry breaking, attributed to fermion flavour mixing \cite{Xue:2016dpl}. This point will be further illustrated later when we discuss the generation of neutrino Dirac masses.

\begin{figure*}[t]
\hskip0.9cm\includegraphics[height=3.40in]{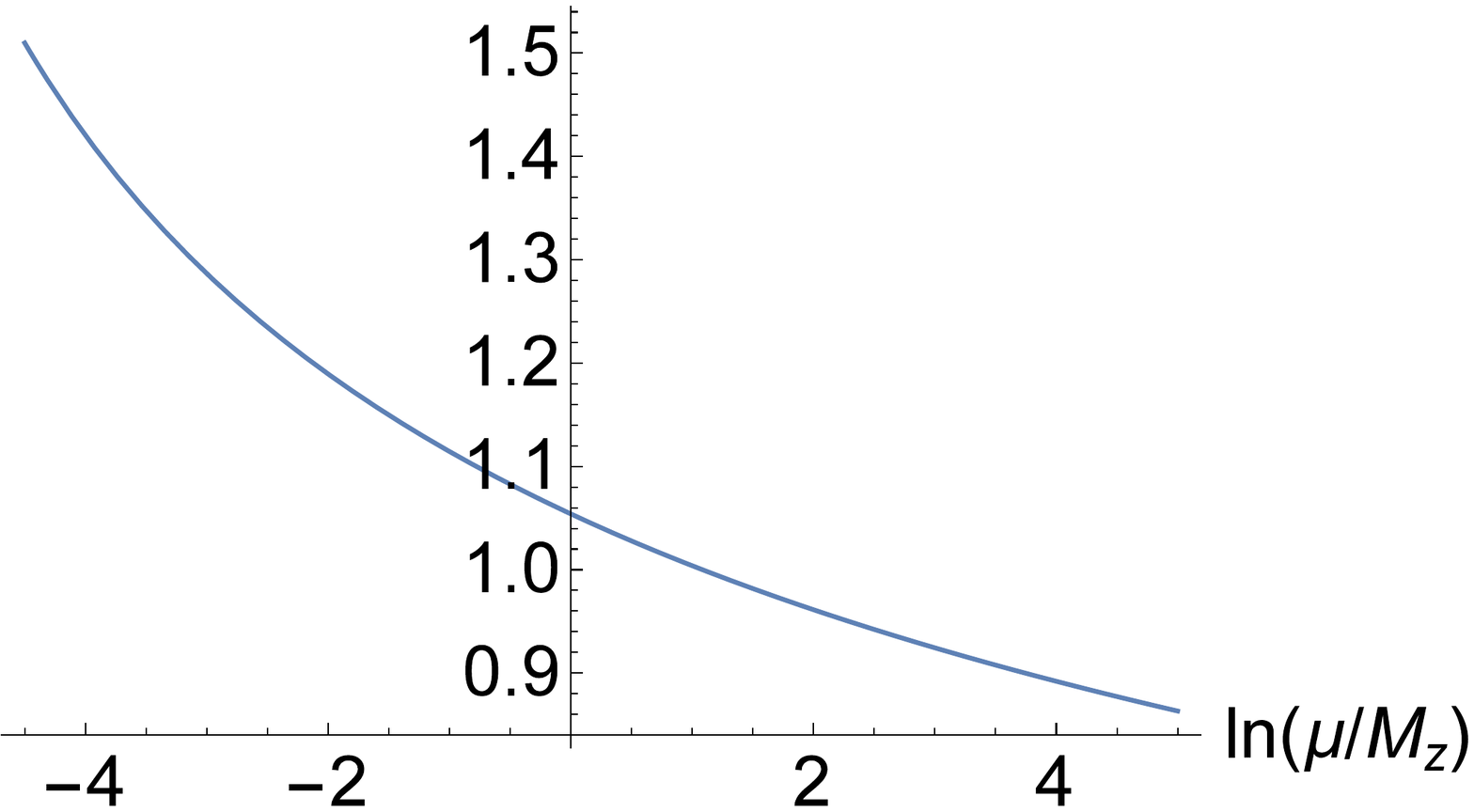}\hskip1.6cm\includegraphics[height=3.40in]{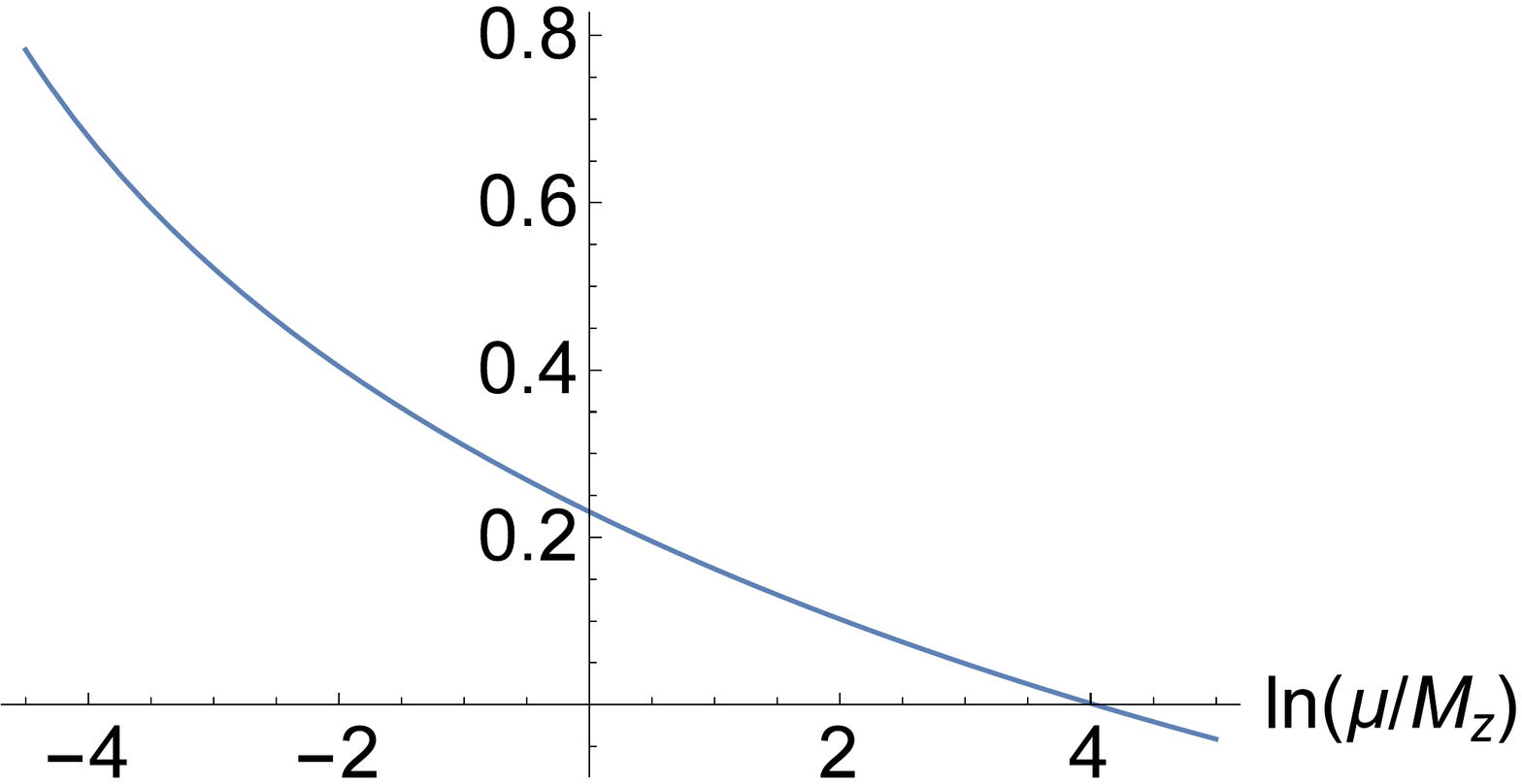}
\put(-115,175){$\tilde\lambda(\mu)$}
\put(-305,175){$\bar g_t(\mu)$}
\vskip-1.1in
\caption{In the top-quark channel, the effective Yukawa coupling $\bar g_t(\mu)$ and quartic couplings 
$\tilde\lambda(\mu)$ as functions of energy scale $\mu$ are determined by RG equations (\ref{reg1},\ref{reg2}), 
mass shell condition  (\ref{thshell}) and experimentally measured top quark and Higgs mass.
The effective quartic coupling $\tilde\lambda(\mu)$ becomes negative at the energy 
scale ${\mathcal E}\approx 5$ TeV.
These figures are reproduced from Refs.~\cite{Xue:2013fla,Xue:2014opa}.} 
\label{figy}
\end{figure*}

\subsection{Experimental values of top-quark and Higgs masses }\label{tope}

To definitely solve the RG equations (\ref{reg1}) and (\ref{reg2}) for 
$\bar g_t$ and $\bar\lambda$, it requires boundary conditions at a physical energy scale. BHL naturally introduced the theoretical compositeness condition $\tilde Z_H(\Lambda_{\rm cut})=1/\bar g_t^2(\Lambda_{\rm cut})=0$
and $\tilde\lambda(\Lambda_{\rm cut})=0$ at the composite scale $\sim\Lambda_{\rm cut}$, where the effective Lagrangian (\ref{eff}) is sewed together with the underlying four-fermion Lagrangian (\ref{bhl}). This is the UV completion of the BHL solution at the cutoff $\Lambda_{\rm cut}$. However, their solution to the RG equations (\ref{reg1}) and (\ref{reg2}) with the compositeness condition cannot reproduce simultaneously correct experimental values of the electroweak scale $v$, the top-quark mass $m_t$, and the Higgs boson mass $m_{_H}$. 
Below, we describe our alternative solution to the problem.

Instead of the BHL compositeness condition, 
we obtained \cite{Xue:2013fla,xue2014} the solution to the RG equations (\ref{reg1}) and (\ref{reg2}) by using the boundary conditions based on the experimental values of top-quark and Higgs-boson masses, $m_t\approx 173$ GeV and $m_{_H}\approx 126$ GeV, via
the mass-shell conditions 
\begin{eqnarray}
m_t(m_t)=\bar g_t^2(m_t)v/\sqrt{2}\approx 173 {\rm GeV},
\quad m_{_H}(m_{_H})=[2\bar \lambda(m_{_H})]^{1/2} v\approx 126 {\rm GeV},
\label{thshell}
\end{eqnarray}
as well as the electroweak scale $v= 246$ GeV determined by the measurement of $W^\pm$ and $Z^0$ boson masses. As a result we find the solutions for $\tilde Z_H(\mu)$ 
and $\bar\lambda(\mu)$, as shown in Fig.~\ref{figy}. In low energies $\mu\gtrsim M_z$, the effective Lagrangian and RG equations (\ref{eff},\ref{boun0},\ref{reg1},\ref{reg2}) with experimental boundary conditions (\ref{thshell}) are equivalent to the low-energy SM Lagrangian and RG equations of elementary top-quark and Higgs fields. Extrapolating them to high energies $\mu\gg M_z$, we find $\tilde Z_H(\mu)\not=0$ is finite, 
the composite Higgs boson is a tightly bound state 
and behaves as an elementary particle, 
and the effective quartic coupling $\tilde\lambda(\mu)$ becomes negative at the energy scale $\sim 5$ TeV. This would imply new physics beyond SM at TeV scales, which will be briefly explained soon.

We will apply the same solution to the sterile neutrino channel (\ref{bhlxl}), studying the spontaneous PQ symmetry breaking of the sterile neutrino sector and its relation to the QCD axion. 
Therefore, it is necessary to briefly explain the UV completion of our solution in contrast with the BHL solution.

The similarities between both solutions are that they approach the scaling domain 
of IR fixed point at low energies ${\mathcal O}(v)$. 
For the energy scale $\mu\gtrsim M_z$, 
$\tilde Z_H(\mu)\not=0$ and $\tilde\lambda(\mu)\not=0$ are finite, 
the composite Higgs boson behaves as an interacting and elementary particle, after the proper 
wave-function (form factor) $\tilde Z_H(\mu)$ renormalisation. The
main differences come up for high energy scale $\mu\gg M_z$. Unlike the BNL solution $\tilde Z_H(\mu)$ and $\tilde\lambda(\mu)$ decrease to zero, our solution $\tilde Z_H(\mu)\not=0$ increases and the effective quartic coupling $\tilde\lambda(\mu)$ becomes negative at the energy scale $\sim 5$ TeV.
This means that the UV completion of our solution is 
drastically different from that of the BHL solution. We give below
a brief summary of our UV completion, more details can 
be found in 
Refs.~\cite{Xue:2014opa,Xue:2016txt,Leonardi:2018jzn}.
\begin{enumerate}[(i)]
\item In the strong four-fermion coupling (\ref{bhl}), there is an SM gauge symmetric phase where composite bosons $\Phi\propto (\bar\psi\psi)$ and fermions 
$\Psi\propto (\bar\psi\psi)\psi$ are formed. The latter can be viewed as a bound state of a composite boson $\Phi$
and an SM fermion $\psi$ \cite{Eichten1986,Xue1996,Xue2000}. They behave as elementary particles as long as their form factors $Z_\Phi$ and 
$Z_\Psi$ do not vanish. 
\item We find \cite{Xue1996,Xue:1999xa} the critical coupling $G_c(\Lambda)$ for the second-order phase transition from the strong-coupling SM symmetric phase ($G>G_c$) to the weak-coupling SM symmetry breaking phase ($G<G_c$). When the running energy $\mu$ becomes smaller than the transition energy scale $\Lambda$ \footnote{In some previous publications, the transition energy scale is indicated by ${\mathcal E}$, while $\Lambda$ stands for the UV cutoff $\Lambda_{\rm cut}$.}, the composite bosons $\Phi$ and fermions $\Psi$ dissolve into SM fermions, as their form factors $Z_\Phi$, 
$Z_\Psi$ and negative binding energies ${\mathcal B}$ vanish. The dissolving dynamics is similar to composite particles (poles) dissolving into their constituents (cuts) in the energy-momentum plane, e.g.~ a deuteron dissolves into a proton and a neutron \cite{Weinberg1963,Weinberg1963a,Weinberg1964,Weinberg1965}. 
\item In the spontaneous SM symmetry breaking phase,  
it is shown \cite{Preparata1996,Xue2013c} that for an energetically favourable ground state, only (Dirac) massive top quark $t$ and composite Higgs boson $\bar t t$ are realised, leading to the BHL top-quark condensate model. Our solution (Fig.~\ref{figy}) implies that the new physics energy scale $\sim 5$ TeV could be the energy scale $\Lambda$ at the transition $G_c$. It is much smaller than the BHL composite scale 
$\Lambda_{\rm cut}$ (\ref{art1}), and the fine-tuning problem is avoided by replacing $\Lambda_{\rm cut}\rightarrow \Lambda$ in the gap equation 
(\ref{tqmassd}). Namely, BHL has not considered the SM gauge symmetric phase (i) of composite particles $\Phi$ and $\Psi$ and the phase transition (ii).
\item The critical point of second-order phase transition plays a role for a fixed point of field theories \cite{Weinberg1976,ZinnJustin2021,Cardy1997,Brezin1976,2016pqf..book.....K,Hooft2017}. The scaling domain of the UV fixed point $G_c(\Lambda)$ renders an SM gauge symmetric effective field theory for composite bosons and fermions at the energy scale $\Lambda$ \cite{Xue:2014opa,Xue:2016txt}, described by relevant RG equations. As the running energy scale, $\mu$ decreases from $\Lambda$ to $v$, the phase transition and dissolving dynamics (ii) proceed, the RG flows take the effective theory of composite particles away from the UV fixed point towards the IR-fixed point of BHL, where an SM gauge symmetry breaking effective theory (\ref{eff}) with a composite Higgs particle is realised. However, it is hard to quantitatively study these properties, because of the non-perturbative nature of strong critical four-fermion coupling.
\item Based on these discussions, we expect there are at least two ``matching'' conditions. First, the compositeness condition at the cutoff scale $\Lambda_{\rm cut}$, where the SM gauge symmetric effective Lagrangian of composite bosons $\Phi$ and fermions 
$\Psi$ in the UV scaling domain is sewed together with the underlying four-fermion Lagrangian (\ref{art1}) for their form factors 
$Z_\Phi=Z_\Psi=0$. Second, the dissolving condition at the energy scale $\Lambda\sim 5$ TeV, where the BHL effective Lagrangian (\ref{eff}) of composite Higgs boson is sewed together with the effective Lagrangian of composite bosons $\Phi$ and fermions $\Psi$ by matching their form factors $Z_\Phi$ and $Z_\Psi$ to those in Eq.~(\ref{boun0}), as well as matching RG flows between UV and IR scaling domains. These are 
non-perturbative issues and will be subjects for future studies. 
\end{enumerate}

\section{Spontaneous breaking of sterile neutrino PQ symmetry}\label{sterile}

The Lagrangian (\ref{bhlxl}) 
with the three right-handed neutrinos $\nu^\ell_R$ preserves the sterile neutrino lepton-number symmetry $U_{\rm lepton}(1)$, which can be also called as the sterile neutrino hypercharge 
symmetry $U_{Y_R^\nu}(1)$. The right-handed neutrinos $\nu^\ell_R$ hypercharge $Y_R^\nu$ is proportional to their $B-L$, 
where $B$ and $L$ are the baryon and lepton numbers of particles.
We identify this global chiral symmetry $U_{\rm lepton}(1)$ 
as the PQ chiral symmetry $U^{\rm PQ}_{\rm lepton}(1)$. 
This means that only sterile neutrinos carry PQ charge $\alpha_{_{\rm PQ}}$, 
$\nu^\ell_{_R}\rightarrow e^{i\alpha_{_{\rm PQ}}}\nu^\ell_{_R}$ 
and $e^{i\alpha_{_{\rm PQ}}}\in U^{\rm  PQ}_{\rm lepton}(1)$.

When three right-handed neutrinos $\nu^\ell_R$ are added into the SM fermion content, the $U_{B-L}(1)$ symmetry is anomaly free and the gauge-gravitational anomaly is also zero, see for example see Ref.~\cite{Akhmedov1993,Ma_2015,Heeck_2014}.  
The hypercharge $U_Y(1)$ triangle anomaly-free requires 
$Y_R^\nu=Y_L+1$, where $Y_L$ is the hypercharge of SM $SU_L(2)$ doublet, $Y_R^\nu$ and $Y_L$ remain unconstrained. This gives the possibility of electrically millicharged left-handed neutrino $\nu^\ell_L$, other SM fermions and dequantized electric charges, attributed to the mixing of SM hypercharge $Y_{\rm SM}$ and $(B-L)$ number $U(1)$-symmetries, see for example Refs.~\cite{Babu1989,Babu1990,Babu1992,Giunti:2014ixa}. 

The top-quark channel operator (\ref{bhl}) 
and sterile-neutrino channel operator (\ref{bhlxl})
have the same structure and coupling $G$. 
The sterile-neutrino channel should undergo the SSB dynamics, 
analogously to the top-quark channel. The spontaneous breaking of the chiral $U^{PQ}_{\rm lepton}(1)$ symmetry leads to the generation of sterile neutrino Majorana masses, axion and massive scalar boson. We present in this section the detailed 
studies and results of spontaneous breaking of global 
$U^{PQ}_{\rm lepton}(1)$ symmetry of sterile neutrino channel, in comparison with the spontaneous breaking of SM gauge symmetry of top-quark channel.

It has been discussed that the spontaneous breaking of the PQ 
chiral $U(1)$ symmetry leads to the generation of sterile neutrino 
Majorana mass via a seesaw mechanism, and the violation of lepton-number symmetry, i.e., $U_{B-L}(1)$ symmetry breaking. In this case, the 
Nambu-Goldstone mode is an Axion or a Majoron, and they are equivalent, see Refs.~\cite{Mohapatra1983,Ma_2017}. In addition, the spontaneous symmetry breaking generates the Majorana mass term $\bar\nu^c_{_R}\nu_{_R}$ requires $Y_\nu^R=0$ and then $Y_L=-1$, which gives the same charge quantization and electrically neutral 
neutrinos $\nu_L^\ell$ as in SM. This is consistent with the violation of the $U_{\rm B-L}(1)$ symmetry by the Majorana mass term, which forbids the mixing of the SM hypercharge and $(B-L)$ number \cite{Babu1989,Babu1990,Giunti:2014ixa}.

\subsection{Sterile neutrino 
condensate and composite bosons}\label{neutrinoSb}

To start this section, let us mention the studies \cite{Martin1991,Antusch2003,Smetana2013}, where the third neutrino family 
is incorporated into the top-quark condensate model. The Majorana mass term 
$M\bar\nu_R^c\nu_R$ is explicitly introduced in Lagrangian, while the Dirac neutrino 
condensate $\langle\bar\nu_L\nu_{R}\rangle$ is spontaneously developed via the 
operator $G\bar\nu^{\ell}_{_L}\nu^{\ell}_{_R}\bar\nu^{\ell}_{_R}\nu^{\ell}_{_L}$ 
(\ref{art1}) in the third SM family $(\ell=\tau)$. It is shown that the 
phenomenological aspects of the top-quark condensate model \cite{Bardeen1990} have 
been improved. Instead, here we study not only the Dirac neutrino condensate, but 
also the Majorana neutrino condensate $\langle\bar\nu^{\ell\, c}_R\nu^{\ell\,}_{R}\rangle$ 
developed by the $G\bar\nu^{\ell c}_{_R}\nu^{\ell}_{_R}\bar\nu^{\ell}_{_R}\nu^{\ell c}_{_R}$ (\ref{art1}), 
leading to the spontaneous generation of Majorana neutrino masses.

Similarly to the $\langle\bar t_{a}t_{a}\rangle$ condensate and top-quark Dirac mass generation,
see Eqs.~(\ref{bhl}) and (\ref{tqmassd}) 
in Sec.~\ref{SSBS}, the four-fermion operator (\ref{bhlxl}) undergoes SSB and develops  
$\langle\bar\nu^{f\, c}_R\nu^{f\,}_{R}\rangle$ condensate and generates the sterile neutrino Majorana mass. The four-fermion coupling $G$ is identical, the family index ``$f$'' and $N_f=3$ play the same role as the color index ``$a$'' and $N_c=3$. Using the approach of large-$N$ expansion, 
as indicated by the tadpole diagram in Fig.\ref{tadpole}, the 
chiral $U^{\rm  PQ}_{\rm lepton}(1)$-symmetry is spontaneously broken 
by non-vanishing vacuum expectation value $m^M\not=0$, 
\begin{eqnarray}
m^M=-G\sum_{f=1,2,3}\langle\bar\nu^{f\, c}_R\nu^{f\,}_{R}\rangle
=2 G\frac{i}{(2\pi)^4}\int^\Lambda d^4p \frac{m^M}{p^2-(m^M)^2}.
\label{mam}
\end{eqnarray}
This generates the Majorana mass $m_3^M=m^M$ of the most massive sterile neutrino $N_R^3$ (mass eigenstate), 
analogously to top-quark mass generation. However,
the sterile neutrino condensate (\ref{mam}) violating the sterile neutrino 
(lepton) number by two units, differently from the top-quark condensate case, where the quark-number conservation is not violated. 

The Nambu-Goldstone theorem guarantees the productions of a sterile massless Nambu-Goldstone boson, i.e.~the pseudoscalar bound 
state 
\begin{eqnarray}
\phi^M=i\sum_{f=1,2,3}\langle\bar\nu^{f,c}_R\gamma_5\nu^f_R\rangle, 
\label{mps}
\end{eqnarray}
and a sterile massive scalar particle, i.e.~the scalar bound state 
\begin{eqnarray}
\phi^M_{_H}=\sum_{f=1,2,3}\langle\bar\nu^{f,c}_R\nu^f_R\rangle. 
\label{mhs}
\end{eqnarray}
Both of them carry two units of the sterile neutrino lepton number. These composite 
bosons $\phi^M$  and $\phi^M_{_H}$ are represented by the poles appearing 
in the bubble sum, as shown in Feynman diagrams of Fig.~\ref{bubblesum}.  

In the same line presented in Ref.~\cite{Bardeen1990}, 
the calculation can be done straightforwardly 
by replacements $t_L\rightarrow \nu^c_R$, 
$t_R\rightarrow \nu_R$, $\bar t_L\rightarrow \bar \nu^c_R$, and $\bar t_R\rightarrow \bar \nu_R$.
The bubble diagram, see the left diagram of Fig.~\ref{tadpole}, is given by
\begin{eqnarray}
\Pi_{s,p}(q^2)&=&i(G/2)\int d^4x e^{iqx} \langle\bar\nu^{f,c}_R\Gamma_{s,p}\nu^f_R(x),\bar\nu^{f,c}_R\Gamma_{s,p}\nu^f_R(0)\rangle_{\rm connected}\nonumber\\
&=& (G/2)\Big[\Pi_{s,p}(\mu^2_{s,p}) + (q^2-\mu^2_{s,p})\Pi^\prime_{s,p}(\mu^2) 
\Big]
\label{bubble0} 
\end{eqnarray}
where $\Gamma_{s}=1, \mu^2_{s}\not=0$ for the scalar channel and 
$\Gamma_{p}=\gamma_5, \mu^2_{p}=0$ for the pseudo scalar channel. The gap equation (\ref{tqmassd}), represented by the tadpole 
diagram in Fig.~\ref{tadpole}, requires $(G/2)\Pi_{s,p}(\mu^2_{s,p})=1$. As a result, the sum 
of bubble diagram Fig.~\ref{bubblesum}, gives the poles of scalar and pseudo scalar propagators,
\begin{eqnarray}
\Gamma_{s,p}(q^2)&=&\frac{(G/2)}{1-(G/2)\Pi_{s,p}(q^2)}=\frac{\Pi^{\prime ^{-1}}_{s,p}(\mu^2)}{(q^2-\mu^2_{s,p})}.
\label{bubble} 
\end{eqnarray}
They represent the pesudoscalar composite boson (\ref{mps}) and the scalar 
composite boson (\ref{mhs}).
The pseudoscalar Nambu-Goldstone boson  
$\phi^M$ is an Axion (or a Majoron). The massive scalar boson is a Higgs-like boson, in analogy to the case of SM gauge symmetries breaking by top-quark condensate. Both composite bosons carry two unite of sterile-neutrino lepton numbers. 

\subsection{Low-energy effective Lagrangian of dark matter particles}\label{neutrinoS}

Analogously to the discussions in Sec.~\ref{SSBS} for the top-quark condensate model, 
the effective Lagrangian of dark matter particles at the low-energy scale $\mu$ is given by
\begin{eqnarray}
L^S_{\rm eff} &=& L^S_{\rm kinetic} + g_{t0}(\bar N^{3,c}_R N^3_R\phi^M_{_H}+ {\rm h.c.})
\nonumber\\ 
&+& Z_\phi|\partial_\mu {\phi^M_{_H}}|^2-m_{\phi}^2{\phi^M_{_H}}^\dagger {\phi^M_{_H}}
-\frac{\lambda_{0}}{2}({\phi^M_{_H}}^\dagger {\phi^M_{_H}})^2\nonumber\\
&+& g_{t0}(\bar N^{3,c}_R \gamma_5 N^3_R\phi^M + {\rm h.c.})+ Z_\phi|\partial_\mu {\phi^M}|^2+ \Delta L^S_{\rm SM},
\label{effs}
\end{eqnarray}
where $L^S_{\rm kinetic}$ is the {\it bilinear} sterile neutrino 
kinetic terms. Analogously to the top quark in Eq.~(\ref{eff}), $N^{3}_R$ indicates the heaviest mass eigenstate of sterile neutrinos in the third lepton family. Apart from the massive scalar boson $\phi^M_{_H}$ 
Lagrangian in analogy with the $\langle \bar tt\rangle$-condensate Lagrangian (\ref{eff}), the massless pseudoscalar boson kinetic term $Z_\phi|\partial_\mu {\phi^M}|^2$ and its interaction with sterile neutrinos are present. The $\Delta L^S_{\rm SM}$ represents a possible effective Lagrangian of $\phi^M_{_H}$ and $\phi^M$ interacting with SM particles. 
The conventional renormalization $Z_{{\nu_{R}}}=1$ for fundamental 
sterile neutrinos and the unconventional wave-function renormalization (form factor)
$\tilde Z_\phi$ for the composite scalar boson $\phi^M_{_H}$ and 
pseudo scalar boson $\phi^M$ are adopted
\begin{equation}
\tilde Z_\phi(\mu)=\frac{1}{\bar g^2_s(\mu)},\, \bar g_s(\mu)=\frac{Z_{\phi Y}}{Z_\phi^{1/2}}g_{t0}; \quad \tilde \lambda_s(\mu)=\frac{\bar\lambda_s(\mu)}{\bar g^4_s(\mu)},\,\bar\lambda_s(\mu)=\frac{Z_{4\phi}}{Z_\phi^2}\lambda_0,
\label{boun0s}
\end{equation}
where $Z_{\phi Y}$ and $Z_{4\phi}$ are proper renormalisation constants of 
the Yukawa coupling and quartic coupling in Eq.~(\ref{effs}). In the IR-domain for SM physics via top-quark condensate, see Sec.~\ref{top}, 
the dark matter particle effective Lagrangian is realised 
at the experimentally unknown sterile scale 
$v_{\rm s}\equiv v_{\rm sterile}$. The full one-loop RG equations for 
running couplings $\bar g_s(\mu^2)$ and $\tilde \lambda_s (\mu^2)$ 
are given by
\begin{eqnarray}
16\pi^2\frac{d\bar g_s}{dt} &=&\frac{9}{2}\bar g_s^3,
\label{reg1s}\\
16\pi^2\frac{d\bar \lambda_s}{dt} &=&12\left[\bar\lambda_s^2+\bar g_s^2\bar\lambda_s -\bar g^4_s \right],\quad t=\ln\mu ,\label{reg2s}
\end{eqnarray}
which are the same as RG equations (\ref{reg1}) and (\ref{reg2}), 
but absence of gauge interactions. 
The SSB-generated sterile neutrino Majorana mass $m^M(\mu)=\bar g_s^2(\mu)v_{\rm s}/\sqrt{2}$. 
The composite scalar boson $\phi^M_{_H}$ is described by its pole-mass 
$m^2_\phi(\mu)=2\bar \lambda_s(\mu) v_{\rm s}^2$, form-factor 
$\tilde Z_\phi(\mu)=1/\bar g_s^2(\mu)$ 
and effective quartic coupling $\tilde\lambda_s(\mu)$, provided that finite wave-function renormalisation (form factor) $\tilde Z_\phi(\mu)>0$ and effective quartic coupling $\tilde\lambda_s(\mu)>0$ are obeyed. 
The sterile neutrino Majorana mass $m^M$ and sterile scalar 
particle mass $m^M_\phi$ satisfy the mass-shell conditions, 
\begin{eqnarray}
m^M=\bar g_s(m^M)v_{\rm s}/\sqrt{2},\quad (m^M_\phi)^2/2
=\bar\lambda_s (m^M_\phi) v_{\rm s}^2,
\label{smass}
\end{eqnarray}
which are the boundary conditions for RG equations (\ref{reg1s}) 
and (\ref{reg2s}). 
The scale $v_{\rm s}$ represents the energy scale of the PQ chiral symmetry 
$U^{\rm PQ}_{\rm lepton}(1)$ breaking and 
lepton-number violation \cite{Xue2015,Xue:2016dpl}. Analogously 
to the RG equations (\ref{reg1}) and (\ref{reg2}), Equations 
(\ref{reg1s}) and (\ref{reg2s}) are RG equations in the IR 
scaling domain \cite{Bardeen1990,Cvetic1999}. 
The boundary conditions (\ref{smass}) for RG equations (\ref{reg1s}) 
and (\ref{reg2s}) of the sterile neutrino sector are similar to the boundary conditions (\ref{thshell}) of the RG equations (\ref{reg1}) and (\ref{reg2}) 
for the top-quark channel, as discussed in Sec.~\ref{tope}.

\begin{figure*}[t]
\hskip1.1cm\includegraphics[height=1.2in]{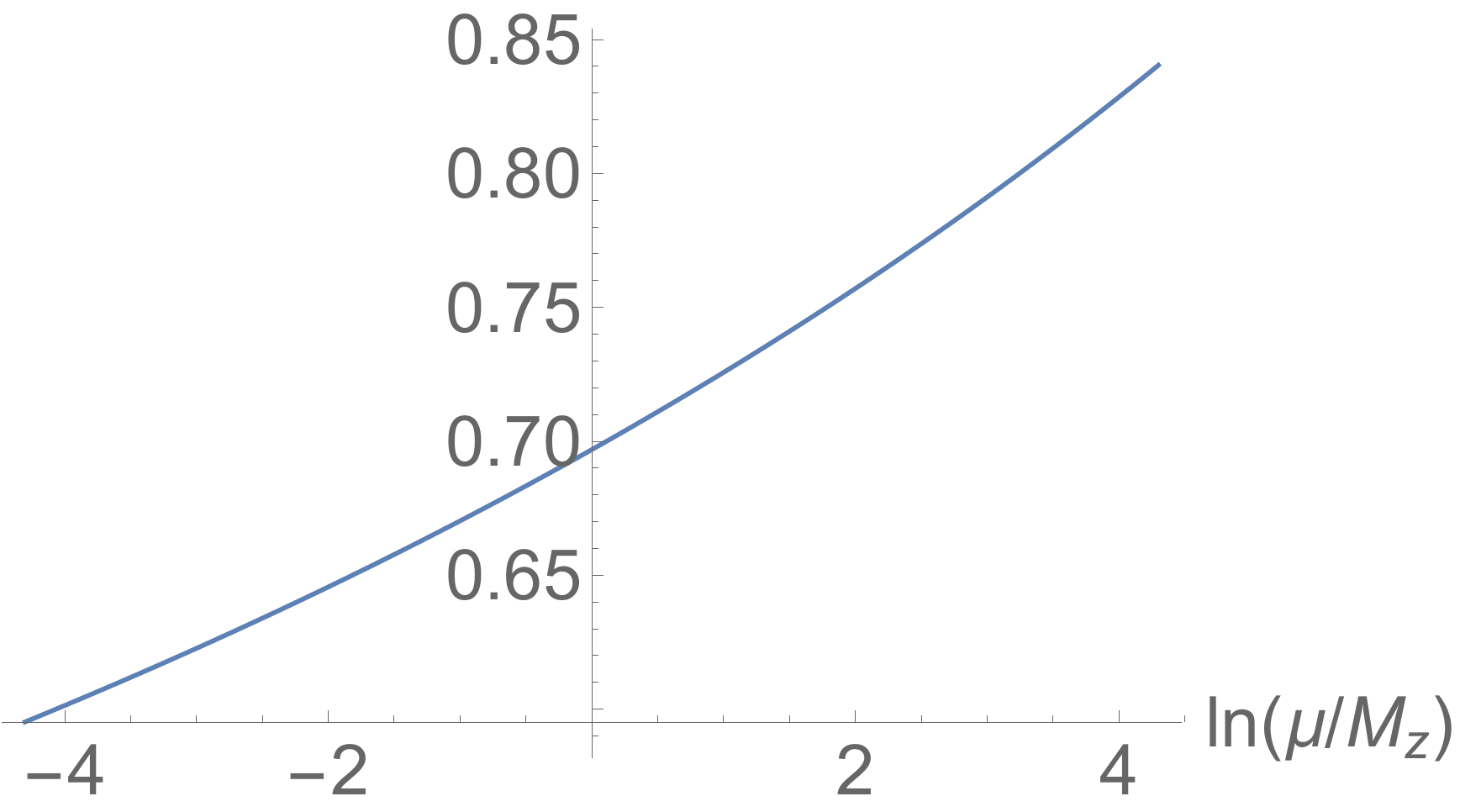}\hskip2.1cm\includegraphics[height=1.2in]{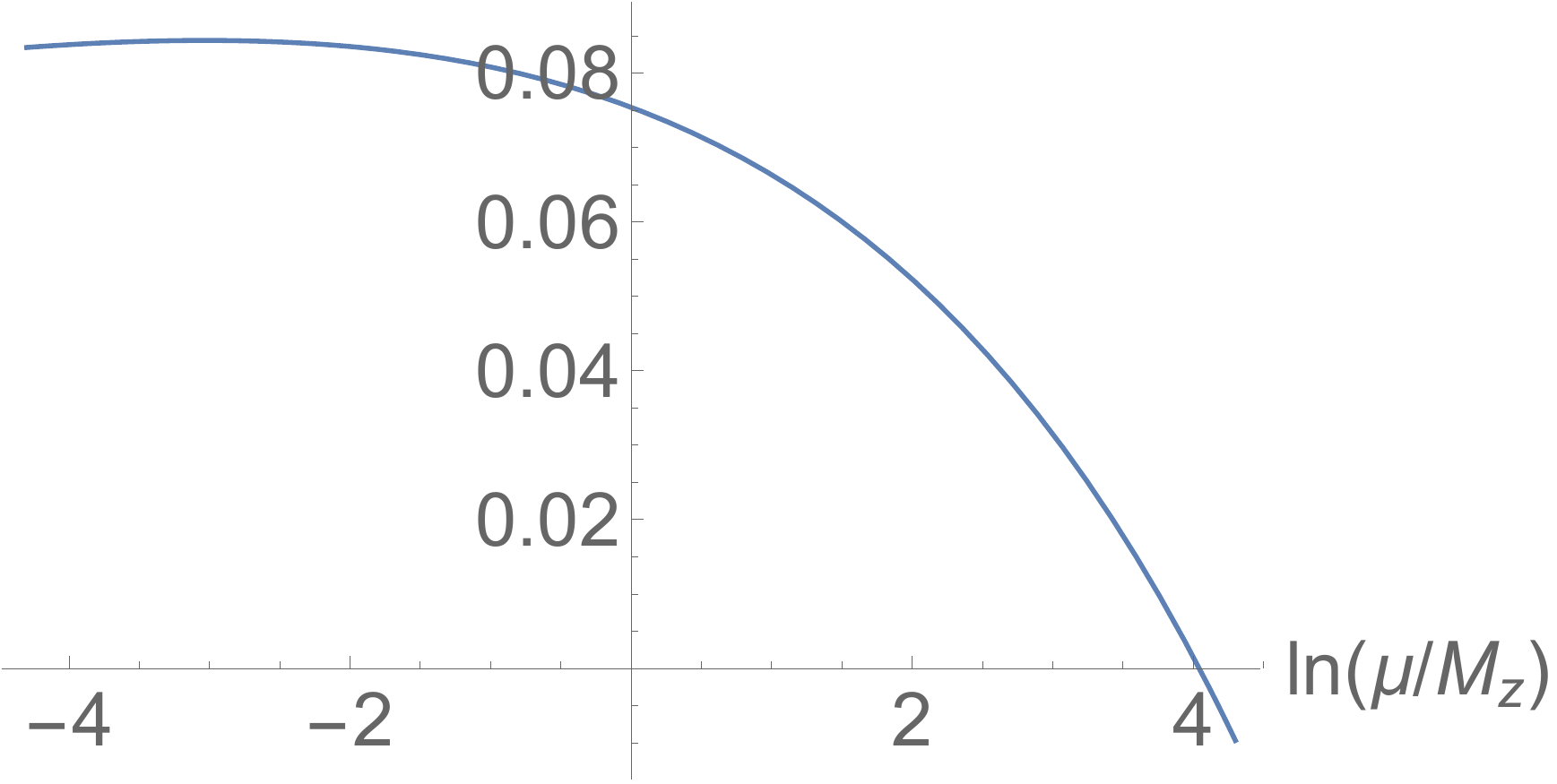}
\put(-115,95){$\tilde\lambda_s(\mu)$}
\put(-305,95){$\bar g_s(\mu)$}
\caption{In the sterile-neutrino channel, the effective Yukawa coupling $\bar g_s(\mu)$ and quartic couplings 
$\tilde\lambda_s(\mu)$ as functions of energy scale $\mu$ are determined by RG equations (\ref{reg1s},\ref{reg2s}), 
mass shell condition (\ref{smass}). We chose the heaviest neutrino mass $m^M\approx 0.83 ~m_t\approx  143$ GeV 
and $m_\chi\approx 0.92 ~m_{_H}\approx  116$ GeV by demanding $\tilde\lambda_s =0$ at the energy scale $\Lambda\approx 5$ TeV, for the reasons in text. At this energy scale,
the quartic coupling vanishes $\tilde\lambda =0$ as that in the top-quark channel (see, the right plot of Fig.~\ref{figy}). \comment{Note that the Yukawa coupling $\bar g_s$ is identified as the third family one 
$\bar g^{(3)}_s$, i.e., $\bar g_s\equiv \bar g^{(3)}_s$, 
see Eq.~(\ref{m3}).}
} \label{figys}
\end{figure*}

In the effective Lagrangian (\ref{effs}), the pseudo scalar boson $\phi^M$ and scalar boson $\phi^M_H$ are tightly bound states of right-handed sterile neutrinos. They behave as elementary bosons, 
since their wave-function renormalization (form factor) 
$\tilde Z_\phi(\mu)=1/\bar g_s^2(\mu)$ is finite.
After proper wave-function renormalization (\ref{boun0s}), the composite 
pseudo scalar boson $\phi^M$ (\ref{mps}) and scalar boson $\phi^M_{_H}$ (\ref{mhs}) are defined as axion and massive $\chi$boson
\begin{eqnarray}
\phi^M\Rightarrow a,\quad  
\phi^M_{_H} \Rightarrow \chi,\quad m_{\phi}\Rightarrow m_{\chi}. 
\label{ax}
\end{eqnarray}
The low-energy effective Lagrangian of dark energy particles is
\begin{eqnarray}
L^S_{\rm eff} &=& L^S_{\rm kinetic} + \bar g_{s}(\bar N^{3 c}_R N^{3}_R \chi+ {\rm h.c.})
\nonumber\\ 
&+&|\partial_\mu {\chi}|^2-m_{\chi}^2{\chi}^\dagger {\chi}
-\frac{\tilde \lambda_{s}}{2}({\chi}^\dagger {\chi})^2\nonumber\\
&+& \bar g_{s}(\bar N^{3 c}_R \gamma_5 N^{3}_R a + {\rm h.c.})+
|\partial_\mu a|^2 + \Delta L^S_{\rm SM}.
\label{effsax}
\end{eqnarray}
Due to the absence of sterile neutrino 
directly coupling to gauge bosons, the pseudo scalar Nambu-Goldstone boson 
$\phi^M$ (\ref{mps}) or axion $a$ (\ref{ax}) does not become the longitudinal mode of a gauge boson. This is different from the occurrence in spontaneous SM gauge symmetry breaking. The 
third family Yukawa coupling $\bar g_{s}$ is of order of unit, 
see Fig.~\ref{figys}. The $\chi$boson mass $m_\chi$, sterile neutrino 
mass spectra and kinetic terms $L^S_{\rm kinetic}$ will be discussed 
just below. We will study in details the term 
$\Delta L^S_{\rm SM}$ for possible $\chi$boson and $a$ axion interactions with SM fermions and gauge bosons in Secs.~\ref{axion} and \ref{Xboson}.

\subsection{Mass scales of sterile neutrinos, axion and 
scalar boson}\label{neutrinoM}

Observe that the the top-quark channel (\ref{bhl}) and  
sterile-neutrino channel (\ref{bhlxl})
have the same coupling $G$ and four-fermion interacting structure. 
Moreover RG equations (\ref{reg1},\ref{reg2}) of top-quark channel 
and (\ref{reg1s},\ref{reg2s}) of sterile neutrino channel 
approach to the same low-energy IR scaling domain, where the SM is realized. Therefore, the $U^{\rm PQ}_{\rm lepton}(1)$ breaking scale $v_s$ 
should be the same order magnitude of electroweak gauge symmetry breaking scale $v$,
\begin{eqnarray}
v_s=f_a\approx v=246 {\rm GeV},
\label{fascale}
\end{eqnarray}
where $f_a$ stands for the axion decay constant, relating to the 
form factors of the axion $a$ and $\chi$boson.
The differences between two channels come from the gauge coupling terms in the RG equations (\ref{reg1}) and (\ref{reg2}). However, we cannot determine the RG solutions $\bar g_s(\mu^2)$ and $\tilde \lambda_s (\mu^2)$ of sterile neutrino channel in the same manner of determining $\bar g_t(\mu^2)$ and $\tilde \lambda (\mu^2)$ in the top-quark channel. Because the boundary conditions (\ref{smass}) of sterile particle masses $m^M$ and $m_\chi$ are experimentally unknown. 

Since gauge coupling terms in RG equations 
are perturbative, we do not expect a large qualitative 
difference between the top-quark channel and sterile-neutrino channel RG solutions in the IR scaling domain at the electroweak scale. 
Therefore, we infer that \cite{Xue2015,Xue:2016dpl}
\begin{enumerate}[(i)]
\item\label{m33} the heaviest sterile neutrino 
that we identify as the third family one $N_R^3$ has a Majorana mass $m^M_3\equiv m^M$ that should be of the same order 
of the top quark mass $m_t\approx 173$ GeV, i.e., 
\begin{eqnarray}
m^M_3\sim 10^2 {\rm GeV}
 \label{m3}
\end{eqnarray}
and its Yukawa coupling 
$\bar g_s\sim {\mathcal O}(1)$, similarly to the top quark and Higgs Yukawa coupling;  
\item the sterile axion $a$ is a massless Nambu-Goldstone boson of spontaneous breaking of PQ symmetry in the sterile neutrino sector; 
\item the sterile $\chi$boson mass should be of the same order of the Higgs mass $m_{_H}\approx 126$ GeV, i.e.,   
\begin{eqnarray}
m_\chi\sim 10^2 {\rm GeV}.
\label{mchi}
\end{eqnarray} 
\end{enumerate}
To obtain the couplings $\bar g_s(\mu^2)$ and
$\tilde \lambda_s(\mu^2)$ of sterile neutrino channel,  
we select quantitatively $m^M\approx 0.825 m_t$ and $m_\chi \approx 0.92 m_{_H}$, so that the effective quartic coupling $\tilde\lambda_s$ vanishes at the new physics scale $\Lambda\approx 5$ TeV. The reasons are 
\begin{enumerate}[(i)]
\item in the right plot of Fig.~\ref{figy} for the top-quark channel, 
the quartic coupling 
$\tilde \lambda(\mu^2)$ vanishes $\tilde\lambda(\Lambda)=0$ at this scale, indicating the domain of UV fixed point for new 
physics \cite{Xue:2014opa,Xue:2016txt,Leonardi:2018jzn};
\item the four-fermion sterile neutrino interaction (\ref{bhlxl}) and top quark interaction (\ref{bhl}) have the same structure and coupling, they should undergo the same dynamics not only in the IR domain at electroweak scale $v$, but also in the UV domain at new physics scale $\Lambda$.  
\end{enumerate}
These arguments infer the heaviest sterile neutrino mass 
$m^M_3\sim m_t$, $\chi$boson mass $m_\chi \sim m_{_H}$ and Yukawa coupling 
$\bar g_s\sim {\mathcal O}(1)$ in connection with top quark and 
Higgs boson masses, and their Yukawa coupling. The $\chi$boson is a tightly bound state of sterile neutrinos $\bar N_R^{3,c}$ and $N_R^{3}$ pair. The inequality $m_\chi < 2m^M_3$ follows, due to the negative binding energy. We consider these as previsions of this approach. In addition, the axion decay constant 
$f_a\sim 10^2$ GeV (\ref{fascale}) is drastically different from $f_a > 10^{11}$ GeV in traditional axion models, where $\chi$boson is absent. 
We will come to these points in due course.
  
To end this section, we give the analytical solution to RG equation (\ref{reg1s}),
\begin{eqnarray}
\bar g_s^2(\mu)=\frac{\bar g_s^2(M_z)}{1-\frac{9}{16\pi^2}\bar g_s^2(M_z)\ln \left(\frac{\mu}{M_z}\right)},
\label{gsmu}
\end{eqnarray}
where $M_z$ is the SM $Z^0$ boson mass. Equation (\ref{gsmu}) 
is similar to the QED running coupling in the IR fixed point domain.

\section{Sterile neutrino and warm dark matter particle}\label{warmdm}

\subsection{Sterile neutrino family mixing}\label{neutrinoX}

The sterile neutrino kinetic term $L^S_{\rm kinetic}$ (\ref{effsax}) consists of the 
Dirac mass $m^{D}_\ell$ and Majorana
mass $m^{M}_\ell$ terms 
\begin{eqnarray}
m^{D}_\ell\bar N^{\ell}_LN^\ell_{R}+ m^{M}_\ell\bar N^{c\ell}_{R}N^{\ell}_{R}+ {\rm h.c.},
\label{lmass}
\end{eqnarray}
where $N^\ell_L$ and $N^\ell_R$ respectively represent the mass eigenstates  of normal SM neutrinos ($\nu^\ell_L$) and sterile neutrinos ($\nu^\ell_R$) 
in the $\ell$-th lepton flavor family. Note that 
$N_R^{1,2,3}=N_R^{e,\mu,\tau}$ indicates sterile neutrino in 
$e$, $\mu$ and $\tau$ lepton family respectively. In terms of lepton mass 
eigenstates $( N^l_L,l_L)$ and $( N^l_R,l_R)$, 
gauge eigenstates $\nu^\ell_{L,R}$ and $\ell_{L,R}$ are expressed as, 
\begin{eqnarray}
\nu^\ell_{L,R} = (U^\nu_{L,R})^{\ell l'}N^{l'}_{L,R},\quad \ell_{L,R}= (U^\ell_{L,R})^{\ell l'} l'_{L,R}
\label{gmtran}
\end{eqnarray}
where $U^{\nu}_{L,R}$ and $U^{\ell}_{L,R}$ are $3\times 3$ unitary matrices in lepton 
family flavor space. The unitary lepton-family mixing matrixes are \cite{Xue:2016dpl},
\begin{eqnarray}
U^{\nu\dagger}_LU^{\ell}_L, &\quad&  
U^{\nu\dagger}_LU^\ell_R,\nonumber\\
U^{\nu\dagger}_RU^\ell_L, &\quad&
U^{\nu\dagger}_RU^{\ell}_R.
\label{mmnul}
\end{eqnarray}
The Pontecorvo-Maki-Nakagawa-Sakata (PMNS) family mixing matrix is 
$[(U^\nu_L)^\dagger U^\ell_L]$. Its counterpart in the sector of 
right-handed leptons 
and neutrinos is the family mixing matrix $[(U^\nu_R)^\dagger U^\ell_R]$. 
The mixing between the normal SM neutrinos and sterile neutrinos is given by $[(U^\nu_R)^\dagger U^\ell_L]$ and $[(U^\nu_L)^\dagger U^\ell_R]$. Their counterparts, namely the quark family mixing and quark-lepton mixing matrices can be found in Ref.~\cite{Xue:2016dpl}. The same as the four-fermion coupling $G$ (\ref{art1}), all mixing matrix elements (\ref{mmnul}) are fundamental parameters in our approach. 
We have not so far been able to explain their origins. 
It could be a rearrangement of the SSB ground state due to gauge interactions and flavour physics. It is interesting and worthwhile to consider the scalar democracy \cite{Hill2019,Hill2019a} to study the origin of family mixing elements (angles) in our model. In terms of these family mixing angles (\ref{mmnul}) and SM fermion mass eigenstates, 
the effective four-fermion operators (\ref{art1}) have different values of effective couplings $G$ for different SM fermion species.

On the other hand, these mixing angles play an important role 
in the generation of other SM fermion Dirac masses, 
except for the top-quark mass. In Refs.~\cite{Xue:2015wha,Xue:2016dpl}, we show in details that it is due to the lepton, quark and quark-lepton family mixings via 
$W^\pm$ effective 1PI operators ${\mathcal G}^W_R$ (\ref{rhc0}), 
the Dyson-Schwinger equations (gap equations) for SM fermion self-energy are 
coupled among SM families,  
\begin{eqnarray}
m^D_{ij} =\int^\Lambda  \frac{d^4p}{(2\pi)^4} {\mathcal K}_{ij}(p) \frac{m^D_{ij}}{p^2-(m^D_{ij})^2}
+ {\mathcal M}^D_{ij}[m^D_t,(U_{L,R})^\dagger U_{L,R}],
\label{mam1d}
\end{eqnarray}
where ${\mathcal M}^D_{ij}[m^D_t,(U_{L,R})^\dagger U_{L,R}]$ is an inhomogeneous 
term and ${\mathcal K}_{ij}(p)$ is the Kernel functions of four-fermion 
interactions \footnote{As example, the first term in Eq.~(\ref{art1}) contains the operator $G\bar\nu^{\ell}_{_L}\nu^{\ell}_{_R}\bar\nu^{\ell}_{_R} \nu^{\ell}_{_L}$ that generates neutrino Dirac 
mass terms.} and/or SM gauge interactions. Owing to the explicit symmetry breaking
terms ${\mathcal M}^D_{ij}[m^D_t,(U_{L,R})^\dagger U_{L,R}]$, these inhomogeneous 
gap equations (\ref{mam1d}) admit nontrivial massive solutions, once the top quark 
mass $m^t_D$ is spontaneously generated.
Such generation of SM fermion Dirac masses is attributed to explicit symmetry breaking without extra Goldstone modes. It is due to the quark-lepton mixing, 
charged leptons and neutrinos acquire their Dirac masses $m^{D}_\ell$. As a result, 
such generated SM fermion Dirac mass spectra are closely related to flavour mixing matrices $U_{L,R}$. In summary, the effective four-fermion coupling strength $G$ 
plays the role of generating the top-quark mass by spontaneous symmetry breaking, 
whereas the family flavour mixing angles play the role of generating other SM fermion masses by explicit symmetry breaking. All these parameters and their
relations should be determined, in connection with different phenomena in SM physics. What we can only explain is that fermion masses and flavour 
mixing angles are closely interconnected in this scenario.

\subsection{Sterile neutrino Majorana masses}

In the bilinear sterile neutrino mass terms (\ref{lmass}), 
the Dirac masses $m^{D}_\ell$ are generated by explicit symmetry breaking due to SM family mixing \cite{Xue:2016dpl}, 
as discussed in Eq.~(\ref{mam1d}). While the Majorana 
masses $m^{M}_\ell$ are originated from the the four-fermion interaction 
(\ref{bhlxl}) undergoing the SSB dynamics together with top-quark condensate, see Sec.~\ref{top}. The SSB renders the generation 
of the most massive sterile mass $m^M_3=m^M$. Namely, the diagonal elements of sterile neutrino Majorana mass matrix are $(0,0,m^M_3)$ 
attributed to the SSB. 

Analogously to the Dirac neutrino 
mass generation discussed in Eq.~(\ref{mam1d}),
other light sterile neutrino masses 
$m^M_1$ and $m^M_2$ are generated by explicitly symmetry breaking 
introduced by the four-fermion interaction (\ref{bhlxl}) induced 1PI operators and family mixing $[(U^\nu_R)^\dagger U^\ell_R]$ (\ref{mmnul}) between light sterile neutrinos $N_R^{1,2}$ and the heaviest sterile neutrino $N_R^{3}$. 
The mass-gap equations of sterile neutrinos $N_R^{1,2}$ are,
\begin{eqnarray}
m^M_{1,2}
=2 G\frac{i}{(2\pi)^4}\int^\Lambda d^4p \frac{m^M_{1,2}}{p^2-(m^M_{1,2})^2}
+ {\mathcal M}_{1,2}[m^M_3,(U^\ell_R)^\dagger U^\nu_R],
\label{mam1}
\end{eqnarray}
in right-handed side the first term represents the 
tadpole diagram in Fig.~\ref{tadpole} and the second term ${\mathcal M}$ represents the explicit symmetry breaking contributions 
from the heaviest sterile neutrino mass $m^M_3$ via the
right-handed lepton family mixing $[(U^\ell_R)^\dagger U^\nu_R]$. 
The mass-gap equations (\ref{mam}) and (\ref{mam1}) are coupled together.
The nontrivial and self-consistent solutions of Majorana masses 
$m^M_{1,2}$ should be functions of heaviest sterile neutrino Majorana 
mass $m^M_3$ and family mixing matrix 
elements. \comment{The effective Yukawa couplings of the first and second family sterile neutrinos to $\chi$boson and $a$ axion are also generated 
\begin{eqnarray}
\bar g^{(1,2)}_{s}(\bar N^{(1,2) c}_R N^{(1,2)}_R \chi
+ \bar N^{(1,2) c}_R \gamma_5 N^{(1,2)}_R a) + {\rm h.c.}
\label{effyuk}
\end{eqnarray}
for the low-energy effective Lagrangian (\ref{effsax}) of dark energy particles.}

The sterile neutrino Majorana masses $(m^M_1,m^M_2,m^M_3)$ 
could be of normal hierarchy structure  
$m^M_1< m^M_2<m^M_3 \sim {\mathcal O}(10^2)$ GeV,
depending on the hierarchy of small off-diagonal elements 
of the matrix mixing $[(U^\ell_R)^\dagger U^\nu_R]$ (\ref{mmnul}).
The situation is similar to how to achieve the hierarchy Dirac mass
spectra of SM massive leptons and quarks: the top quark acquires its mass from SSB and other fermions acquire their masses from explicit chiral symmetry breaking induced by the SM family mixing \cite{Xue:2016dpl,Xue:2015wha}. 
The detailed studies of sterile neutrino mass spectra 
will be a future subject, since this is not the scope of this article, 
and we have no experimental information for the mixing matrix 
$[(U^\ell_R)^\dagger U^\nu_R]$ and sterile neutrino mass spectra 
$(m^M_1,m^M_2,m^M_3)$.
 
Nevertheless, we mention the following two points. 
Why only one sterile neutrino Majorana mass is generated by SSB, other two sterile neutrino Majorana masses are generated by explicit symmetry breaking. The reason is that only one Nambu-Goldstone boson (axion) is an energetically favourable configuration of the SSB vacuum ground state of nontrivial Majorana mass. 
This is the same as the reason why only the heaviest top quark Dirac mass is generated by the SSB with only three Nambu-Goldstone bosons becoming the longitudinal modes of SM massive gauge bosons $W^\pm$ and $Z^0$ \cite{Preparata1996,XUE1996583,Xue2013c}.
In addition, employing the sea-saw mechanism 
of the type-I \cite{Minkowski:1977sc,Glashow:1979nm,GellMann:1980vs,Schechter:1980gr,PhysRevLett.44.912}, we 
obtain \cite{Xue:2016dpl} the normal SM neutrinos are 
Majorana and their masses, consistently with
current experiments and observations.  

\section{Effective right-handed electroweak interactions}\label{effR}

In previous section, we adopt the four sterile 
right-handed neutrino $\nu^{\ell\,}_{R}$ operator 
(\ref{bhlxl}) to discuss the spontaneous breaking of sterile neutrino PQ symmetry $U_{\rm lepton}(1)$ and Majorana mass generation, accompanying with sterile pseudoscalar boson 
$\phi^M$ (\ref{mps}) and sterile scalar boson 
$\phi^M_H$ (\ref{mhs}). Such dynamics does not develop the effective electroweak couplings of the sterile neutrino $\nu^{\ell\,}_{R}$, since it is a singlet under all SM gauge groups. 
However, the effective electroweak couplings of the sterile neutrino $\nu^{\ell\,}_{R}$ can be induced by the sterile neutrino $\nu^{f}_{_R}$ and SM fermions $\psi^{f}_{_R}$ 
coupling operator 
\begin{eqnarray}
G\bar\nu^{fc}_{_R}\psi^{f}_{_R}\bar\psi^{f}_{_R} \nu^{fc}_{_R},
\label{bhlxlc}
\end{eqnarray} 
coming from the second term in the effective Lagrangian (\ref{art1}), 
since SM fermions $\psi^{f}_{_R}$ couple to SM gauge bosons. This is the analogy of the effective electromagnetic coupling of SM neutrino $\nu^\ell_L$ developed via Fermi four-fermion interactions $G_F(\bar\nu_L\gamma^\nu\ell)( \bar\ell\gamma_\nu\nu_L)$ 
and $G_F(\bar\nu_L\gamma^\nu\nu)(\bar\ell\gamma_\nu\ell)$ 
between electrically neutral neutrinos $\nu^\ell_L$ and charged leptons $\ell$, which are mediated by massive gauge bosons $W^\pm$ and $Z^0$.

\begin{figure}
\centering
\includegraphics[width=2.2in]{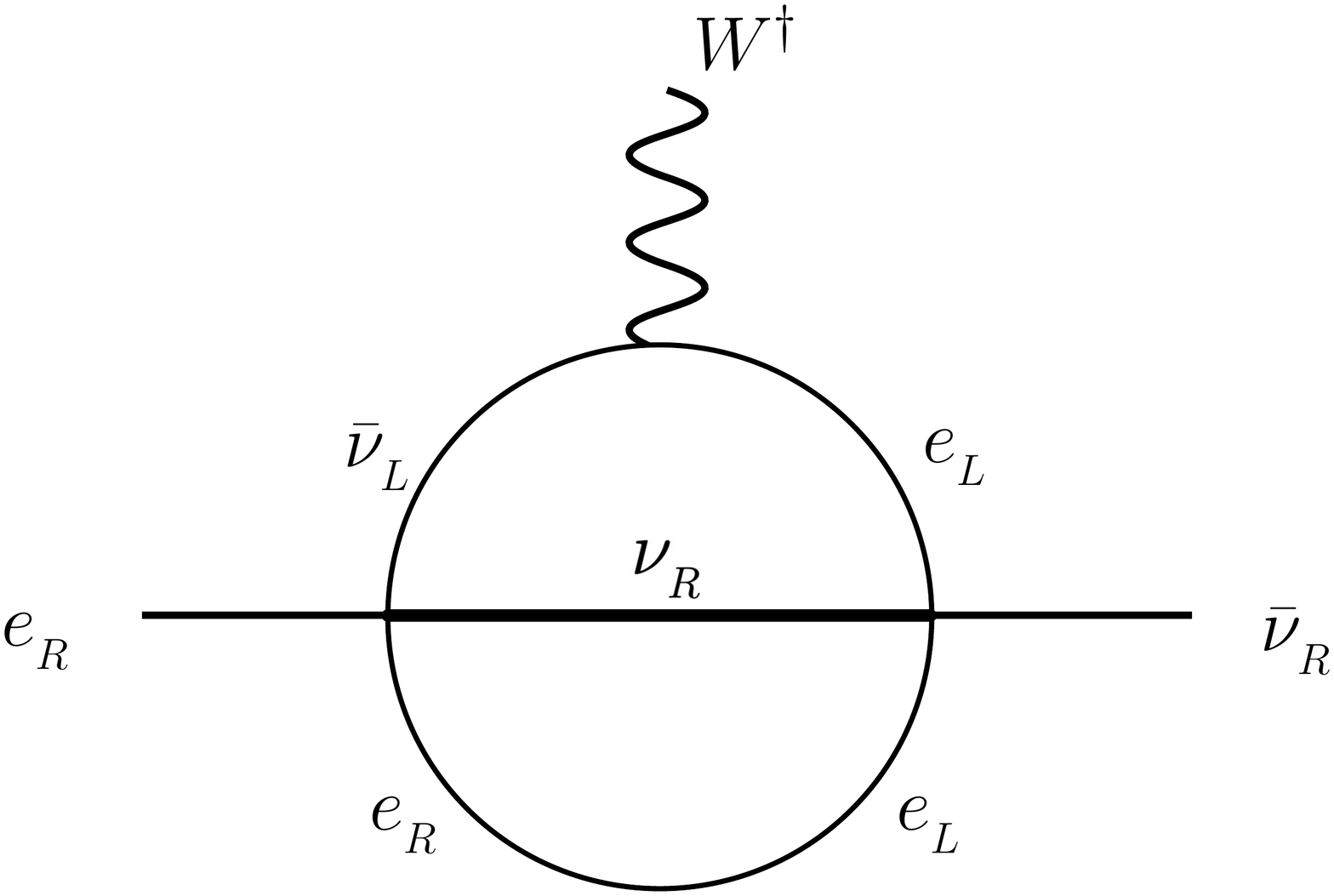}\hskip1.5cm  \includegraphics[width=2.2in]{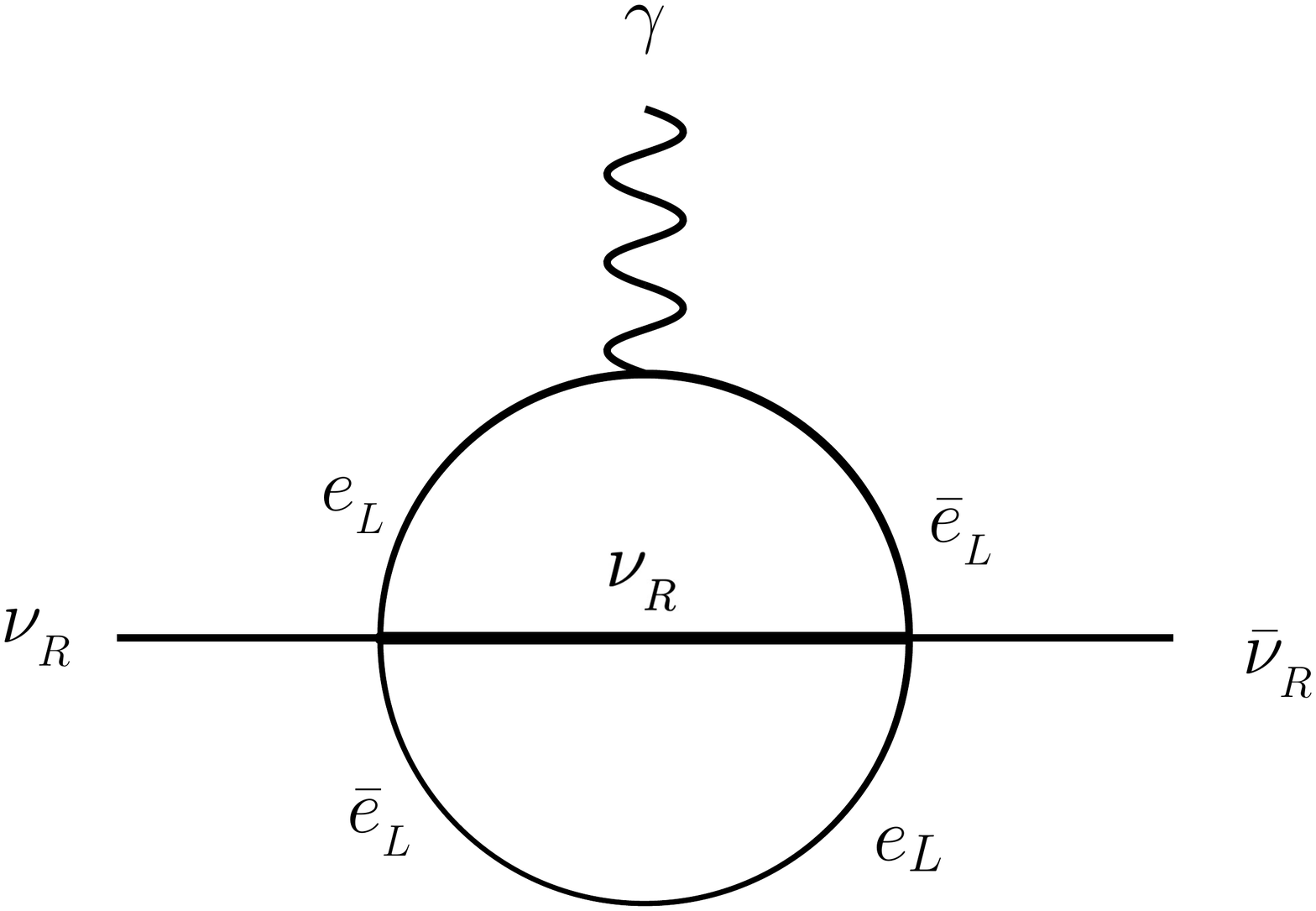}
\caption{This is a sketch to show the possible sunset diagrams 
from the second term of (\ref{art1}), namely, sterile neutrino and SM fermion four-fermion interactions $\bar\nu^{fc}_{_R}\psi^{f}_{_R}\bar\psi^{f}_{_R} \nu^{fc}_{_R}$, here $\psi^{f}_{_R}$ represents SM right-handed fermions. Therefore, via SM fermions, these 1PI vertexes lead to effective SM gauge boson couplings to right-handed neutrinos (\ref{rhc0}). 
Left: the effective 
1PI interacting vertex (\ref{rhc0}) of the gauge boson $W^+$ and right-handed 
sterile neutrino $\nu^\ell_R$, for more details see Figure 3 of Ref.~\cite{Xue:2015wha}. 
Right: the effective 1PI interacting vertex (\ref{rhc0}) 
of photon $\gamma$ and right-handed sterile neutrino $\nu^\ell_R$, and similar one 
for $Z^0$ boson. The slightly thick solid lines inside sunset diagrams 
represent right-handed neutrino propagators with Dirac mass 
(left) or Majorana mass (right). 
A Dirac mass term is present in the internal electron propagator 
from $e_L$ to $e_R$ in the left sunset diagram.}
\label{Rcouplingf}
\end{figure}

\subsection{Induced right-handed neutrino 1PI couplings to SM gauge bosons}

Although the detailed formation of the sterile neutrinos and SM fermions four-fermion interactions (\ref{bhlxlc}) is unknown, in this section 
we discuss qualitatively and phenomenologically 
four-fermion interaction (\ref{art1}) induced 1PI vertexes of sterile neutrinos 
$\nu_R^\ell$ interacting with SM gauge bosons \cite{Xue:1997tz,Xue:1996fm,Xue:2001he} \footnote{There are counterparts of these interactions in the quark sector.}, 
\begin{eqnarray}
	\mathcal{L}&\supset & {\mathcal{G}^W_R}~({g_w}/{\sqrt{2}})\bar \ell_R\gamma^\mu\nu^\ell_RW^-_\mu 
+{\mathcal{G}^Z_R}~({g_w}/{\sqrt{2}})\bar \nu^\ell_R\gamma^\mu\nu^\ell_R Z^0_\mu\nonumber\\ 
	&+& {\mathcal{G}^\gamma_R}~(e)\bar \nu^\ell_R\gamma^\mu\nu^\ell_R A_\mu	+ {\rm h.c.}\,
\label{rhc0}
\end{eqnarray}
where the $SU_L(2)$ gauge coupling $g_w=e/\sin\theta_W$ is defined by the electric charge $e$ and Weinberg angle $\theta_W$ and relates to 
the Fermi constant $G_{F}/\sqrt{2}=g_{w}^{2}/8M_{W}^{2}$ and 
the $W^\pm$ gauge boson mass $M_{W}$. 
These 1PI right-handed interacting vertexes can be induced, for example, from sunset diagrams in Fig.~\ref{Rcouplingf}. 
In the way of coupling parametrization (\ref{rhc0}), 
the effective right-handed couplings 
${\mathcal{G}^W_R}$, ${\mathcal{G}^Z_R}$ and ${\mathcal{G}^\gamma_R}$ are dimensionless. These effective electroweak couplings are functions of the energy scale,
the four-fermion coupling $G$ (\ref{art1}) and flavour mixing 
angles (\ref{mmnul}). Their values should be different, and small enough so as not to violate current experiments and observations at low energies. So far we have not been able to theoretically determine or constrain them. 
Nevertheless, at this preliminary step of modelling, we introduce only one effective coupling ${\mathcal G}_R$
\begin{eqnarray}
{\mathcal G}_R={\mathcal{G}^W_R}\approx {\mathcal{G}^Z_R}\approx {\mathcal{G}^\gamma_R}\ll 1\,
\label{gR0}
\end{eqnarray} 
in the effective right-handed interacting Lagrangian (\ref{rhc0}). The upper limits of these effective couplings must be constrained by Earth-based experiments, astrophysical and cosmological observations.

In Ref.~\cite{Haghighat:2019rht}, using the vertex 
${\mathcal{G}^W_R}({g_w}/{\sqrt{2}})\bar \ell_R\gamma^\mu\nu^\ell_RW^-_\mu$ (\ref{rhc0}), we calculated its contribution to $W^\pm$ boson decay. Comparing with the total $W^\pm$ boson decay width observed, we 
obtain the constraint of coupling $\mathcal{G}^W_R < 4.2\times 10^{-2}$. 
In Ref.~\cite{Shakeri2020}, we calculated 
the sterile neutrino decay rates ($N_R^\ell \rightarrow \nu_L^\ell +\gamma$) 
and obtain the more stringent constraint on the upper limit 
of $\mathcal{G}^W_R< 3.8\times 10^{-4}$, for which sterile neutrinos can be 
candidate of dark matter particles, and satisfy astrophysical constrains. 
Further studies are required to see whether such a 
right-handed electroweak interaction survives the experimental crosschecks.

Moreover, the effective vertexes ${\mathcal{G}^Z_R}({g_w}/{\sqrt{2}})\bar \nu^\ell_R\gamma^\mu\nu^\ell_R Z^0_\mu$ and 
${\mathcal{G}^\gamma_R}(e)\bar \nu^\ell_R\gamma^\mu\nu^\ell_R A_\mu$ (\ref{rhc0}) have the same structure of SM gauge interacting vertexes $({g_w}/{\sqrt{2}})\bar \nu^\ell_L\gamma^\mu\nu^\ell_L Z^0_\mu$ and $(e)\bar\ell\gamma^\mu\ell A_\mu$, where $\nu^\ell_L$ are left-handed neutrinos and $\ell$ stands for charged leptons. This implies a very tiny effective electroweak charge of sterile neutrinos. 
For a given electroweak process, up to the leading order 
at tree level, additional contributions from these 
right-handed electroweak operators (\ref{rhc0}) 
should be at least $(\mathcal{G}^Z_R)^2$ or $(\mathcal{G}^\gamma_R)^2$ times smaller than SM contributions. 
The natural question is whether such effective right-handed electroweak
interactions are consistent with precision experiments, astrophysical and 
cosmological observations, that constrain the upper limits 
of these effective couplings. We are proceeding with these studies.
Here, as an example, we estimate the leading order $(\mathcal{G}^\gamma_R)^2$ 
correction from sterile neutrino $\nu^\ell_{_R}$ bubble diagrams that contribut 
to the photon vacuum polarisation,
\begin{eqnarray}
\alpha\approx \alpha_{\rm sm}\left(1+(\mathcal{G}^\gamma_R)^2\frac{\alpha_{\rm sm}}{15\pi}\frac{m_e^2}{(m^e_N)^2}\right).
\label{alpha}
\end{eqnarray}
Here $\alpha_{\rm sm}$ stands for the SM fine-structure value 
on the electron mass $m_e$ shell, and $m^e_N\approx 90 $ keV 
is the lightest sterile neutrino mass, as will be discussed in the next section \ref{Xenon1T}. The recent precision measurement of fine-structure constant $\alpha$, namely 
$(\alpha_{\rm sm}/\alpha_{\rm exp}-1)\times 10^9 \lesssim - 0.6$ (Figure 1) \cite{Parker2018}, constrains the 
$(\mathcal{G}^\gamma_R)$ upper limit, 
\begin{eqnarray}
\mathcal{G}^\gamma_R< 3.8\times 10^{-4}.
\label{ggamma}
\end{eqnarray}
A more quantitative investigation will be presented elsewhere. In this article, we present a detailed study on the QCD-Axion and two-photon coupling $g_{a\gamma}$, 
whose value depends on ${\mathcal{G}^\gamma_R}$. One of the reasons is that many ongoing experiments and observations are measuring the coupling $g_{a\gamma}$. 
Thus we further have an interconnected constrain on the effective coupling 
${\mathcal{G}^\gamma_R}$ to see any inconsistency.

Before going on to the QCD axion theories and experiments, in order 
to gain more insight into the order of magnitude of effective coupling 
${\mathcal{G}^\gamma_R}\approx {\mathcal{G}_R}$ value (\ref{gR0}) and sterile 
neutrino masses $m^\ell_N$, we briefly recall our study on explaining the 
recent Xenon1T experiment results based on this model (\ref{rhc0}).

\subsection{Xenon1T experiment and sterile neutrinos}\label{Xenon1T}

In the recent article \cite{Shakeri2020}, using the effective interactions (\ref{rhc0}) to account for the Xenon1T experiment results \cite{Aprile2020a},
we find the dominant contribution stemming from 
the 1PI vertex of SM neutrino and sterile neutrino interaction 
in the electromagnetic (EM) channel,
\begin{equation}
(U^\nu_L U^\ell_L)^{l l'} \bar\nu_L^l\Lambda^\mu_{l'} N_R^{l'} A_\mu + {\rm h.c.}.
\label{effem0}
\end{equation} 
It origins from the effective interactions (\ref{rhc0}) and a one-loop Feynman diagram, see Fig.~2 and 4 
of Ref.~\cite{Shakeri2020}. Here $A_\mu$ is the electromagnetic field and $(U^\nu_L U^\ell_L)$ is the PMNS mixing matrix. In the momentum 
space of incoming sterile neutrino $N_\ell(p_{1}^{\mu})$ and outgoing SM neutrinos $\nu_\ell(k_{1}^{\mu})$, 
see the left of Fig.~\ref{nnupf}, the 1PI vertex $\Lambda_\mu$ is given by 
\begin{equation}
\Lambda_{l'}^\mu(q) =i\frac{e g_w^2 {\mathcal G}_Rm_{l'}}{16\pi^2}\Big[(C_0+2C_1)p_1^\mu+(C_0+2C_2)k_1^\mu\Big]\,,
\label{effem1}
\end{equation}
here $m^{l'}$ indicates SM lepton mass.
The coefficients $C_0$, $C_1$ and $C_2$ are the three-point Passarino-Veltman functions \cite{Passarino:1978jh}, computed by the {\tt Package-X} program \cite{Patel:2015tea}. In the low-energy limit 
$q^{2}=(k_1-p_1)^{2}\rightarrow 0$, $C_{0,1,2}\propto M_W^{-2}$. 

The effective operator in Eq.~(\ref{effem0}) represents a peculiar electromagnetic property of normal neutrino and sterile neutrino coupling to photon, stemming from the effective right-handed current coupling in Eq.~(\ref{rhc0}). It is different from the effective electromagnetic operator $\langle\bar\nu_j|J_\mu^{\rm em}|\nu_i \rangle$ of normal neutrinos states $\nu_j$ and $\nu_i$ (diagonal $i=j$ or transition $i\not= j$), e.g., the diagonal neutrino electric form factor $f_Q\bar\nu\gamma_\mu\nu A^\mu$ and magnetic moment $f_M\bar\nu\sigma_{\mu\nu}\nu F^{\mu\nu}$, see review \cite{Giunti:2014ixa}. Instead, the operator (\ref{effem1}) is an effective electromagnetic vertex of normal neutrino state $\nu_L$ and sterile neutrino state $N_R$. It associates the effective neutrino Dirac mass operator $\bar\nu_L N_R$ by the Ward-Takahashi identity \cite{Xue:2016dpl}. We will study its effects on the normal neutrino magnetic moment, electric form factor and charge radius. The effective operator (\ref{effem0}) is more similar to the effective transition magnetic 
moment operator $\mu_{\rm eff} \bar\nu_L\sigma_{\mu\nu}N_R F^{\mu\nu}$, which has been 
intensively discussed \cite{Karmakar:2020rbi,Shoemaker:2020kji,Miranda:2020kwy,Brdar:2020quo}. In the section 5 of 
Ref.~\cite{Shakeri2020}, we compare these two effective operators and obtain the relation 
$\frac{\mu_{\text{eff}}}{\mu_{B}}\sim \frac{G_{F}m_e}{4\sqrt{2}\pi^{2}}
 \mathcal{G}_{R} m_{\tau}$ between the effective coupling 
$\mathcal{G}_{R}$ and the effective transition magnetic moment $\mu_{\text{eff}}/\mu_{B}$. 
As a result, we obtain the constraint on the upper limit of $\mathcal{G}_{R}$ or 
$\mu_{\text{eff}}/\mu_{B}$ from the available experimental and observational data. 
However, we find the most stringent constraint on $\mathcal{G}_{R}$ upper limit comes from the following explanation of the recent Xenon1T data.  

The effective electromagnetic 1PI interacting vertex (\ref{effem0}) or (\ref{effem1}) 
mainly accounts for the Xenon1T experimental result \cite{Aprile2020a} via sterile neutrino $N_R^\ell$ 
inelastic scattering off an electrons bound by nucleus, 
we find the following possible situations \cite{Shakeri2020}:
\begin{enumerate}[(a)]
\item 
Only $N^e_R$ is present today as dark matter component and its Majorana 
mass $m^e_{N}=m^M_1\sim 90$ keV, 
and $N^\mu_R$ and $N^\tau_R$ have already decayed to SM particles. This is the case if
\begin{equation}
\mathcal{G}_R\sim {\mathcal O}(10^{-4})\,;
\label{a}
\end{equation}

\item Sterile neutrinos $N^e_R$ and $N^\mu_R$ are present today as dark matter particles. $N^\tau_R$ has already decayed to the SM particles. 
This indicates $m^\mu_{N}=m^M_2\sim 90$ keV. This is the case if 
\begin{equation}
\mathcal{G}_R\sim {\mathcal O}(10^{-6})\,;
\label{b}
\end{equation}

\item  All sterile neutrinos $N^e_R, N^\mu_R$, and $N^\tau_R$ are present today as dark matter particles. 
This indicates $m^\tau_{N}=m_3^M\sim 90$ keV. This is the case if 
\begin{equation}
\mathcal{G}_R\sim {\mathcal O}(10^{-7})\,.
\label{c}
\end{equation}
\end{enumerate}
To determine which situation is the physical reality, more relevant experiments, 
observations and theoretical studies are still needed. Observe that 
the the situation (a) is consistent with the estimated upper limit (\ref{ggamma}) from 
the $\alpha$ precision measurement \cite{Parker2018} and assumption (\ref{gR0}). 
This implies that the situation (a) could be the most possible case. 
Nevertheless, it is sure that the absolutely upper limit of $\mathcal{G}_R$ coupling is at least,
\begin{equation}
\mathcal{G}^\gamma_R\sim\mathcal{G}_R <10^{-4},
\label{gcp}
\end{equation}
for the effective right-handed electroweak interactions (\ref{rhc0}). 

We further speculate the situation (a) 
$m^e_{N}=m^M_1\sim 10^2$ keV, theoretically inferred $m^\tau_{N}=m^M_3\sim 10^2$ GeV (\ref{m3}),
and $m^\mu_{N}=m^M_2\sim 10^2$ MeV. This $m^\mu_{N}$ value 
is inferred by assuming $N^\mu_R$ mediating the process leading to events observed in the MiniBooNE experiment \cite{Bertuzzo2018,Gninenko:2009ks,Magill:2018jla}. 
However, to confirm the neutrino $N_R^e$ as viable warm 
dark matter particle, one still needs to study not only their properties 
consistently constrained by all cosmological and astrophysical 
observations, but also possible direct and/or indirect detections 
in laboratory experiments.

To end this section, we have to mention that the induced 1PI
EM vertexes (\ref{rhc0}) and (\ref{effem1}) possibly explain anomalies or predict new effects due to: (i) sterile neutrinos $N_R^e,N_R^\mu$ and $N_R^\tau$ 
produced by SM neutrinos $\nu^e_L,\nu^\mu_L$ and $\nu^\tau_L$ 
quasi-elastic scattering off a nucleus; (ii) sterile neutrinos $N_R^e,N_R^\mu$ 
and $N_R^\tau$ produced by photons and their annihilation to photons. We are proceeding 
the studies on the anomalous muon-magnetic moment \cite{Abi2021}, and MiniBooNE and LSND 
experiments \cite{Bertuzzo2018,Gninenko:2009ks,Magill:2018jla}. 
In next section, we will study the QCD axion physics by using the 1PI vertex 
${\mathcal{G}^\gamma_R}~(e)\bar \nu^\ell_R\gamma^\mu\nu^\ell_R A_\mu$ (\ref{rhc0}), 
and the constrain ${\mathcal{G}^\gamma_R}\lesssim 10^{-4}$ (\ref{ggamma}) or (\ref{gcp}),
as well as sterile neutrino masses $m^\ell_{N}$.

\comment{
In addition, we also discuss there how sterile neutrino relic 
abundance \cite{Allahverdi:2020bys,Boyarsky:2018tvu} 
is possibly consistent with the current dark matter 
relic density obtained from CMB observations  
$\Omega_{DM}h^{2}=0.120\pm 0.001$ \cite{Aghanim2020}. 
Moreover, considering sterile neutrinos may influence the extra number of effective neutrinos $\Delta N_{\text{eff}}$ decoupled from the thermal bath in the early Universe \cite{Drewes:2013gca}. The parameter $\Delta N_{\text{eff}}$ is constrained by CMB \cite{Aghanim2020}. Its exact value in our scenario also depends on the cosmic history before the big bang nucleosynthesis and after the inflationary era~\cite{Gelmini:2019esj,Drewes:2013gca} which is unknown then we will not consider its computation here.}

\section{\bf Sterile QCD axion and superlight dark matter particle}\label{axion}

We have discussed the possible axion candidate, which is a pseudoscalar bound state of sterile neutrino and anti sterile neutrino pair. 
It is a Nambu-Goldstone boson of the broken $U^{\rm PQ}_{\rm lepton}$ symmetry. The $U^{\rm PQ}_{\rm lepton}$  symmetry associates with sterile neutrinos only.  
As discussed in Sec.~\ref{neutrinoM}, the symmetry breaking scale 
$v_s$ is the same order of the electroweak scale 
$v\approx 246$ GeV, and the axion decay constant (form factor) 
$f_a\approx v_s$. Henceforth, the scale relation 
$v_s=f_a\approx v$ (\ref{fascale}) is imposed. 
In this section, we show that this sterile axion essentially plays 
the role of PQ QCD axion,   
solving the strong CP problem in QCD. 

\subsection{Peccei-Quinn axion approach to strong CP problem}\label{QCDaxion}

First, we briefly recall the original PQ axion model. 
The SM should possess a global chiral PQ $U(1)$ 
symmetry \cite{Peccei1977,Peccei1977a}, which is necessarily spontaneously broken with a Nambu-Goldstone axion \cite{Weinberg1978,Wilczek1978}. Under the PQ transformation, the axion field translates as 
$a(x)\rightarrow a(x)+ \alpha_{_{\rm PQ}} f_a $. 
On the other hand, the PQ current has a chiral anomaly,
\begin{eqnarray}
\partial_\mu J^\mu_{PQ} =\delta {\mathcal L}^{A}_{\tilde g\tilde g}/\delta a= \xi\frac{ g^2_s}{32\pi^2
}F_{\mu\nu}^a\tilde F^{\mu\nu}_a + g_{a\gamma}\frac{e^2}{32\pi^2}F_{\mu\nu}\tilde F^{\mu\nu},
\label{canomaly} 
\end{eqnarray} 
and the corresponding effective Lagrangian
\begin{eqnarray}
{\mathcal L}^{A}_{\tilde g\tilde g} = \xi\frac{a}{f_a}\frac{ g^2_s}{32\pi^2
}F_{\mu\nu}^a\tilde F^{\mu\nu}_a + g_{a\gamma}\frac{a}{f_a}\frac{e^2}{32\pi^2}F_{\mu\nu}\tilde F^{\mu\nu},
\label{lanomaly} 
\end{eqnarray}
where $g_s$ is the $SU_c(3)$ strong coupling, 
$e$ is the electric charge  ($\alpha=e^2/(4\pi)$), $\xi$ is a dimensionless 
coefficient and $g_{a\gamma}$ is the coupling of axion and two photons. 
The anomalous term (\ref{lanomaly}) adds to the 
QCD Lagrangian with the static CP-violation term 
${\mathcal L}^{\bar\theta}_{QCD}=\bar \theta \frac{ g^2_s}{32\pi^2}F_{\mu\nu}^a\tilde F^{\mu\nu}_a$. The CP invariant QCD vacuum demands the vacuum expectation value $\langle F_{\mu\nu}^a\tilde F^{\mu\nu}_a\rangle \equiv 0$ at the minimum $\langle a\rangle=-\bar \theta f_a/\xi$ of axion field potential 
${\mathcal L}^{\theta}_{QCD} + {\mathcal L}^{A}_{\tilde g\tilde g}$.  
Upon this minimum, the physical axion field is then represented by the massive fluctuation field $a-\langle a\rangle$. This is the PQ dynamical solution to the strong CP problem of QCD. The QCD axion coefficient $\xi$ value depends on axion models,
for more detailed discussions, see review \cite{Peccei2006}.

In the original PQ QCD axion model, the PQ charge is associated to 
the SM quarks, two Higgs fields are introduced to make the SM 
invariant under PQ $U(1)$ transformation \cite{Peccei1977,Peccei1977a,Weinberg1978,Wilczek1978,1978PhLB...74..229B,Bardeen1978,BARDEEN1987401}. 
In this model, the spontaneous breaking of PQ $U(1)$ symmetry is achieved by nonvanishing Higgs field vev $f_a$ and the coefficient $\xi$ is identified
\begin{eqnarray}
f_a=v=246 {\rm GeV}\quad {\rm and }\quad \xi\sim {\mathcal O}(1) 
\label{pqxi0} 
\end{eqnarray}
in the effective Lagrangian (\ref{lanomaly}). 
The axion-photon coupling $g_{a\gamma}$ and the axion mass $m_a$ are given by, 
\begin{eqnarray}
g_{a\gamma}=2\xi\frac{ m_u}{m_u+m_d},\quad m_a=\xi m_\pi \frac{f_\pi}{f_a}\frac{\sqrt{m_um_d}}{m_u+m_d},
\label{amass} 
\end{eqnarray}
where $m_{u,d}$ are $u,d$ quark masses ($m_u/m_d\approx 0.46$), 
$m_\pi\approx 135$ MeV and $f_\pi\approx 93$ MeV are pion 
mass and decay constant. The PQ axion model (\ref{pqxi0}) has no any free parameter, in the sense that both $f_a$ and $\xi$ are fixed. However, 
such low-energy axion model 
has been ruled out experimentally, since the model yields the branching ratio \cite{BARDEEN1987401}
\begin{eqnarray}
Br(K^+\rightarrow \pi^+ +a)\approx 3\times 10^{-5}\xi^2,
\label{bran} 
\end{eqnarray}
which is well above the KEK bound \cite{Asano1981} 
$Br(K^+\rightarrow \pi^+ +{\rm nothing})< 3.8\times 10^{-8}$. This implies 
that for the scale $f_a=v$, 
the dimensionless coefficient $\xi < 3.56\times 10^{-2}$ at least.

\begin{figure}[t]
\begin{center}
\includegraphics[width=0.49\hsize]{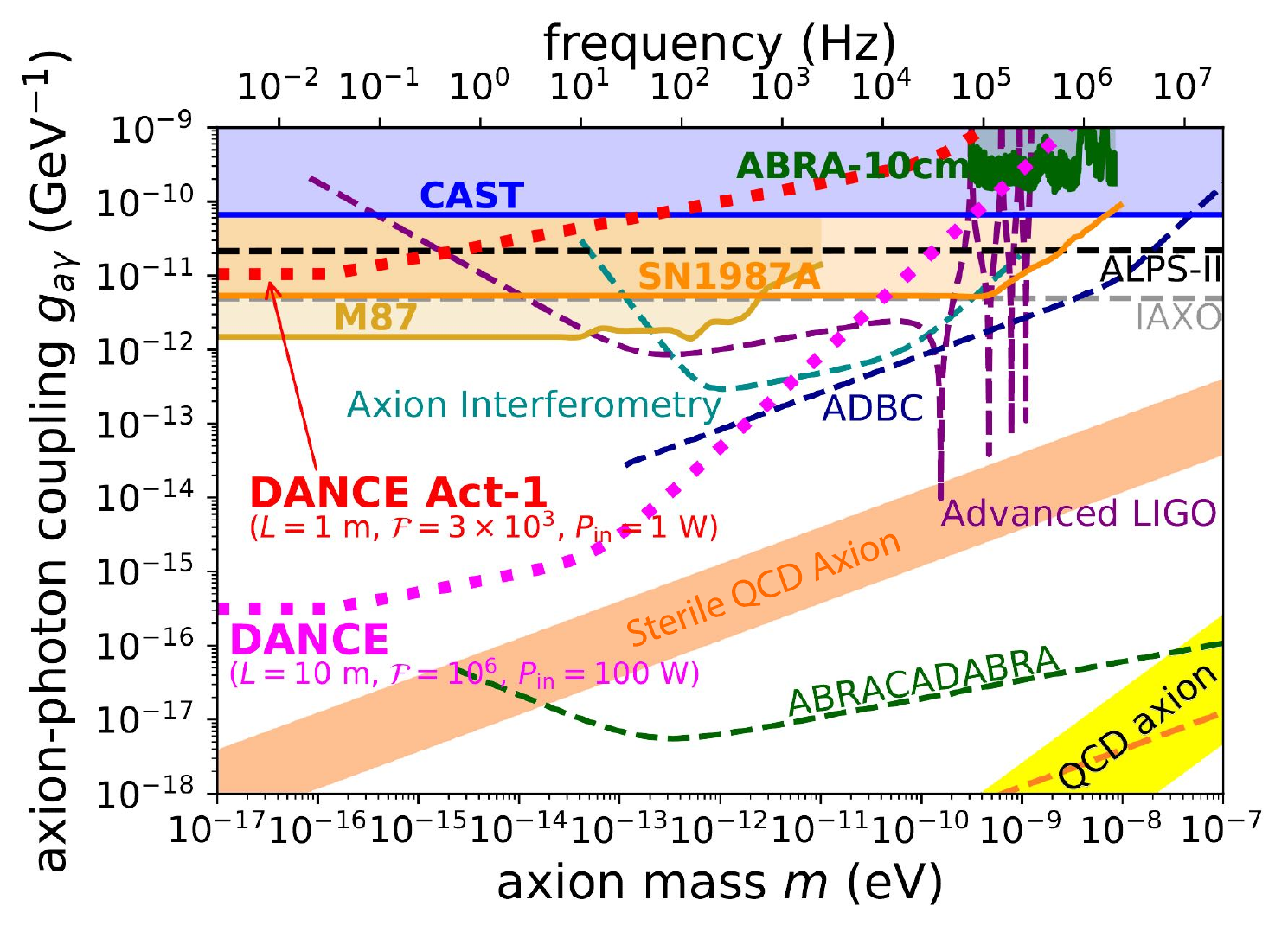}
\includegraphics[width=0.50\hsize]{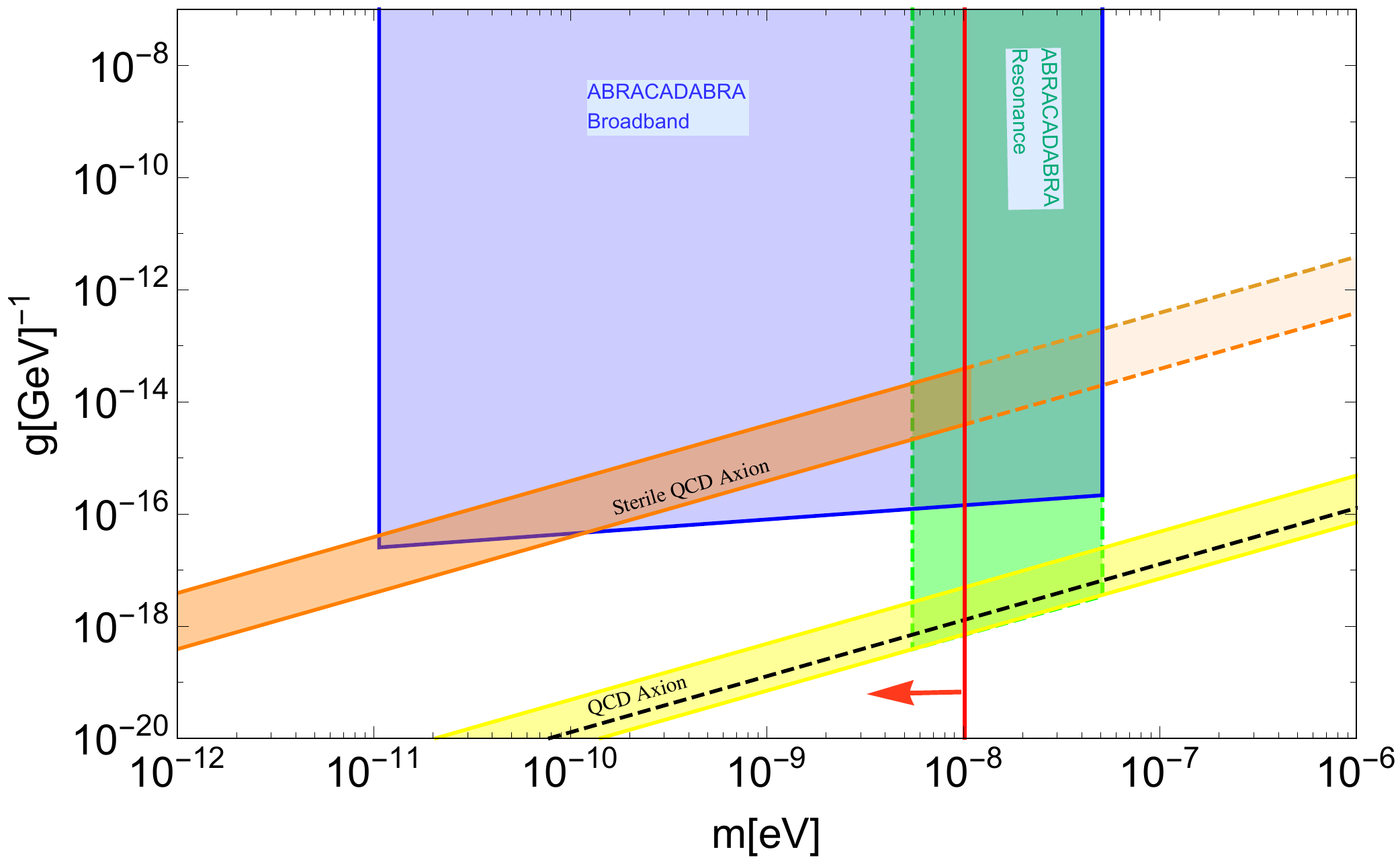}
\caption{The Left: The sensitivity curves for the axion-photon 
coupling $g^{\rm s,exp}_{a\gamma}$ in terms of axion mass $m_a$. 
The left figure is produced from the figure 2 of Ref.~\cite{Michimura2020}.
The green dashed line below is the proposed sensitivity 
to be reached by the upgraded ABRACADABRA experiment. The sterile QCD axion model result (\ref{magag}) is plotted and indicated. 
The PQ relation (\ref{amgag}) is plotted (dashed line), located within the QCD axion yellow region. The Right: The sterile QCD axion model result (\ref{magag}) with the absolutely upper limits (\ref{apexp}) 
(vertical red line) can be probably probed by the proposed ABRACADABRA broadband configuration with 5~T and 1~m$^3$~\cite{Kahn2016}.
\comment{The expected shot noise limited sensitivity of DANCE and DANCE Act-1 with a year-long integration time are shown as dotted lines. The solid lines with the shaded region are current bounds obtained from CAST~\cite{CAST} and ABRACADABRA-10cm~\cite{ABRA-10cm} experiments, and the astrophysical constraints from the gamma-ray observations of SN1987A~\cite{SN1987A} and the X-ray observations of M87 galaxy~\cite{M87}. The dashed lines are projected limits from ALPS-II~\cite{ALPS-II}, IAXO~\cite{IAXO} and ABRACADABRA broadband configuration with 5~T and 1~m$^3$~\cite{Kahn2016}. The expected limits from Axion Interferometry proposal with intra-cavity power of 1~MW~\cite{AxionInterferometry}, the expected integrated limits from ADBC experiment with intra-cavity power of 1~MW~\cite{ADBC}, and the expected sensitivity of Advanced LIGO with a scheme proposed in Ref.~\cite{NaganoAxion} are also shown with dashed lines for comparison.}
} 
\label{Bounds}  
\end{center}
\end{figure}

In order to see the possible ways for solving this problem, we rewrite the PQ relation (\ref{amass}) as
\begin{eqnarray}
g^{\rm exp}_{a\gamma} \equiv g_{a\gamma}\frac{\alpha}{2\pi f_a}&=&\frac{\alpha m_a}{\pi f_\pi m_\pi}\left(\frac{ m_u}{m_d}\right)^{1/2},\label{amgag}\\
g^{\rm exp}_{a\gamma}  ({\rm GeV})^{-1} 
&=& 1.26\times 10^{-10} m_a  ({\rm eV}),\label{amgag1}
\end{eqnarray}
and the modeling coefficient $\xi$ relates to the scale $f_a$
\begin{eqnarray}
\xi =g^{\rm exp}_{a\gamma} \left(\frac{m_u+m_d}{m_u}\right) \frac{\pi f_a}{\alpha}.
\label{xilimit} 
\end{eqnarray}
The PQ linear relation (\ref{amgag}), which is independent of $f_a$, locates (dashed line) in the yellow region of QCD axion in Fig.~\ref{Bounds}.  
The relation (\ref{xilimit}) implies that if the scale 
$f_a$ is a free parameter $f^{\rm inv}_a \gg v$, 
the $\xi\sim {\mathcal O}(1)$ possibly agrees with 
the small experimental value $g^{\rm exp}_{a\gamma}$. 
This leads to the one-parameter QCD axion models
at high energies.

High-energy QCD axion models introduce new quark fields which carry PQ charge but are the SM gauge singlets. Their vev scales are much larger than the electroweak scale, i.e., $f^{\rm inv}_a\gg v$. This is the essential difference 
between the original PQ low-energy axion model and its variants of high-energy 
axion models. Basically, two types of high-energy axion models have been proposed. The KSVZ model due to Kim \cite{Kim1979} and 
Shifman, Vainshtein and Zakharov \cite{Shifman1980} introduces a scale field 
$\sigma$ with $f^{\rm inv}_a=\langle\sigma\rangle\gg v$ 
and associates PQ charge only to a super-heavy quark of mass $M_Q\sim f^{\rm inv}_a$. 
The DFSZ model, due to Dine, Fischler and Srednicki \cite{Dine1981} and Zhitnisky \cite{Zhitnitsky1980}, 
adds to the original PQ model an SM singlet scalar field $\phi$ 
which carries PQ charge and $f^{\rm inv}_a=\langle\phi\rangle\gg v$. In Ref.~\cite{Achiman1991}, the vacuum expectation value $\langle\phi\rangle\not= 0$ is achieved by the sterile neutrino condensation. 
The KSVZ and DFSZ linear relations of coupling $g^{\rm exp}_{a\gamma}$ and mass $m_a$ locate in the yellow QCD axion region of Fig.~\ref{Bounds}. 
The spread of the yellow QCD axion region depends on the detailed modelling 
of dimensionless coefficient $\xi$ values in the effective 
Lagrangian (\ref{lanomaly}). Their essential difference from the original 
PQ model is that the free scale $f^{\rm inv}_a\gg v$ is chosen value so as to achieve small $g^{\rm exp}_{a\gamma}$ and $m_a$ values, and meet the constraints presented in Fig.~\ref{Bounds}. 
These two models are also called invisible models because the effective interactions (\ref{lanomaly}) 
between the axion and SM particles  are very small for 
$f^{\rm inv}_a > 10^{11}$ GeV.

In the next section, we show how the QCD axion can be realized at the electroweak scale $f_a=v$, consistently with experimental and 
observational constraints, in terms of small $\xi$ value 
due to the very tiny coupling of photon and sterile neutrino.

\subsection{Sterile-neutrino QCD axion model}\label{nuaxion}

We associate PQ charge only to sterile (right-handed) neutrino $\nu_{_R}$ which is an SM gauge singlet. The PQ chiral symmetry $U^{\rm PQ}_{\rm lepton}$ is spontaneously broken by the four-sterile-neutrino interaction 
(\ref{bhlxl}). Sterile neutrinos acquire Majorana masses, accompanying with a composite Nambu-Goldstone axion $a$ and composite scalar 
$\chi$boson, which carry PQ charges (sterile neutrino numbers). 
The symmetry breaking scale $f_a\approx v$ represents the composite scale of the composite axion $a$ and $\chi$boson and their decay constants. We show now how these composite bosons couple to SM particles and PQ chiral anomalies associating to QED and QCD gauge fields are produced.  

\begin{figure}
\centering
\includegraphics[width=1.8in]{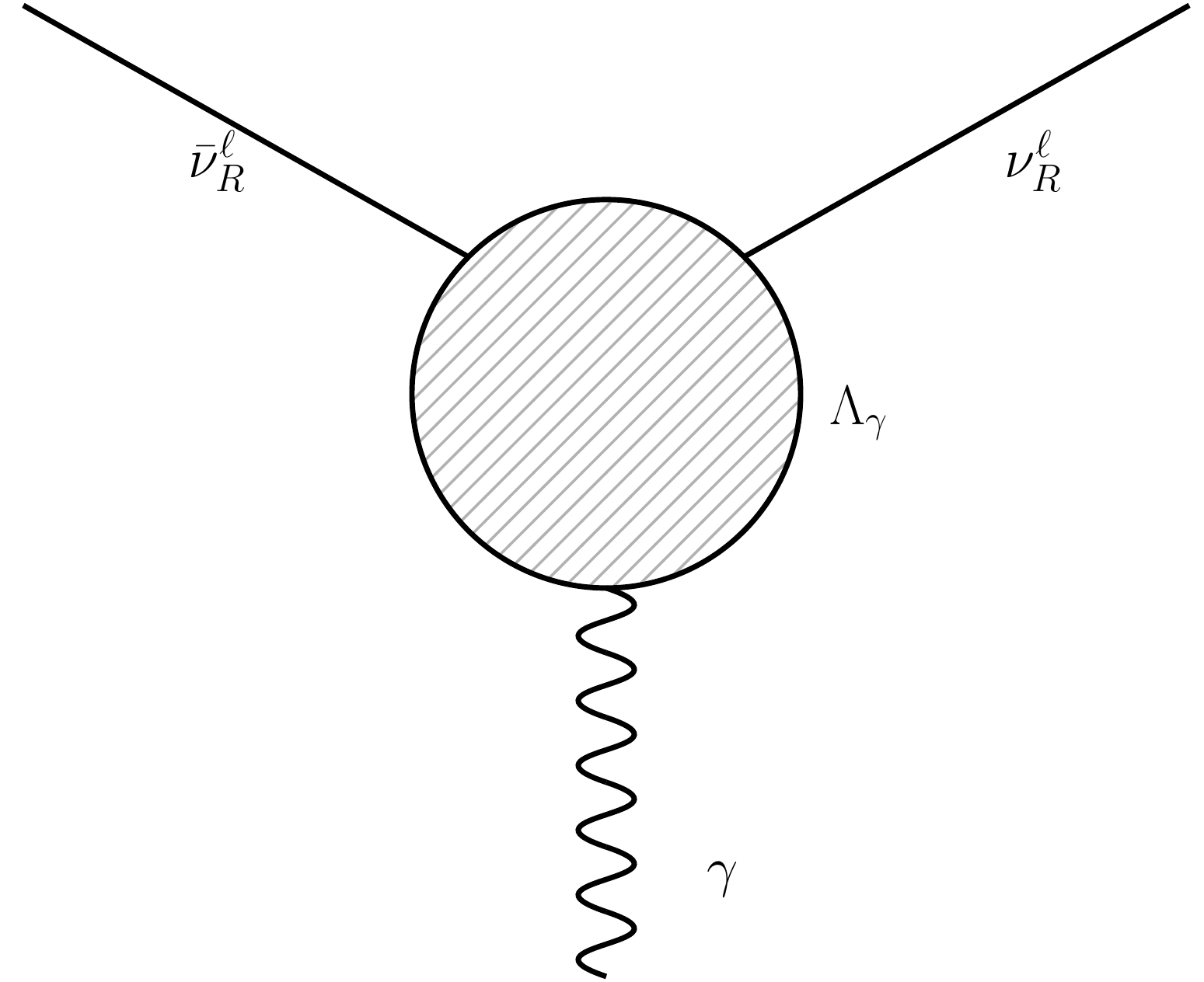}\hskip1.5cm  \includegraphics[width=3in]{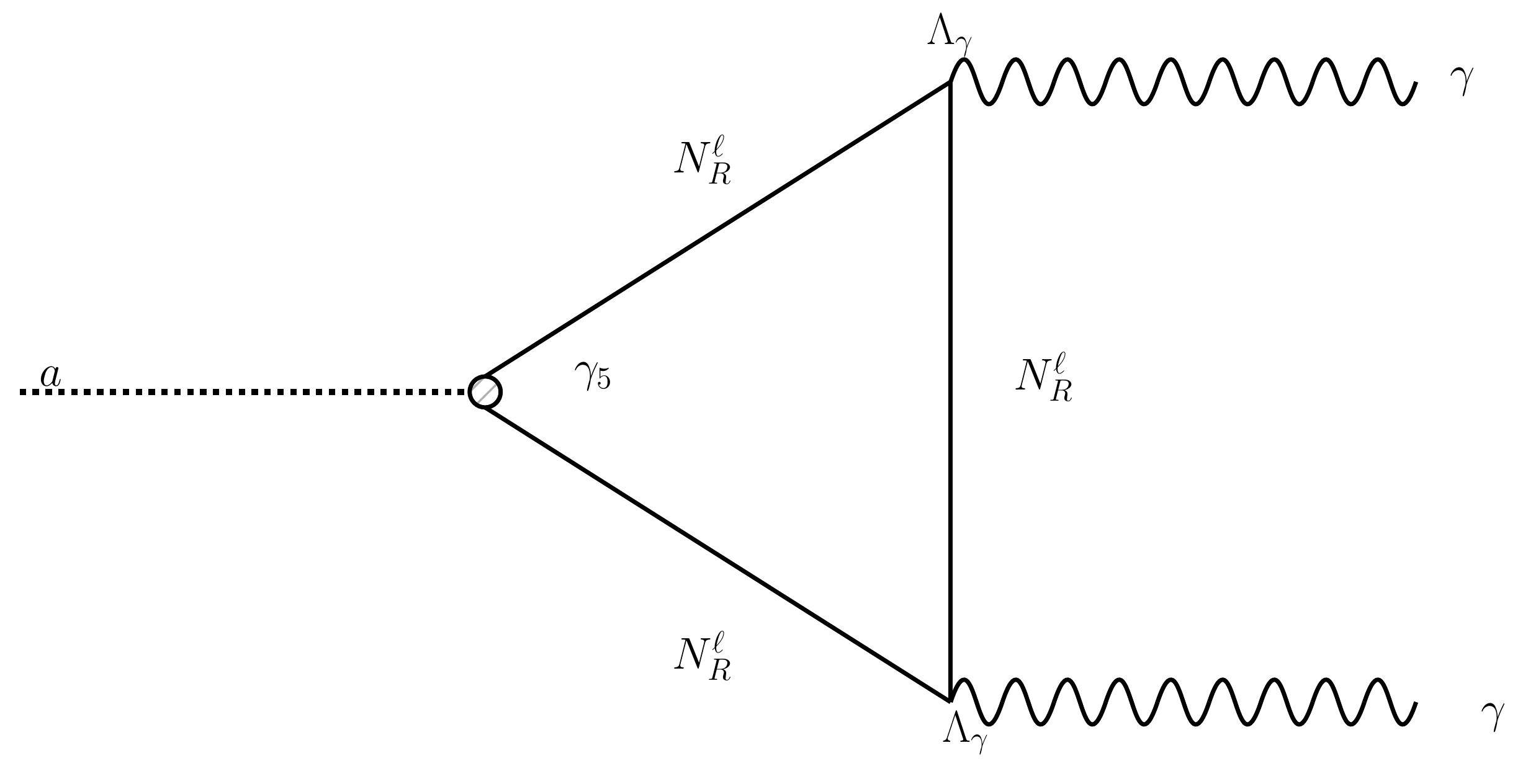}
\caption{Left: the effective 
interacting vertex (\ref{rhc0}) 
of photon $\gamma$ and right-handed sterile neutrino $\nu^\ell_R$: 
$\Lambda_\gamma\propto e{\mathcal G}_R^\gamma\bar \nu^\ell_R\gamma^\mu\nu^\ell_R A_\mu=
e {\mathcal G}_R^\gamma\bar N^\ell_R\gamma^\mu N^\ell_R A_\mu$. 
Right: the triangle diagram of massive sterile neutrino $N_R^\ell$ loop with 
one axial Yukawa coupling vertex $(m^M_3/f_a)\gamma_5 $ to an axion (dot line), 
and two coupling vertexes $\Lambda_\gamma$ to two photons (wave lines). 
The contribution to the triangle diagram is dominated by the heaviest sterile neutrino $N^3_R$ channel. Note that the solid lines in the triangle loop represent massive sterile neutrino 
$N_R^\ell$ propagators, in which the Majorana mass term
$m^{M}_\ell\bar N^{c\ell}_{R}N^{\ell}_{R}$ 
(\ref{lmass}) is present.}
\label{nnupf}
\end{figure}

\subsubsection{axion coupling to two photons and SM fermions}\label{axionSM}


Based on the photon channel vertex (\ref{rhc0}) and triangle diagram 
of Fig.~\ref{nnupf}, we use the standard approach to calculate triangle anomaly, and obtain an anomalous 1PI contribution to the effective Lagrangian at low energies,  
\begin{eqnarray}
{\mathcal L}_{\rm eff} &\supset& g^s_{a\gamma}\frac{a}{f_a}\frac{e^2}{32\pi^2}F_{\mu\nu}\tilde F^{\mu\nu},
\label{panomaly} 
\end{eqnarray}
where $f_a=v$ at electroweak scale (\ref{fascale}). This anomalous 1PI contribution (\ref{panomaly}) yields the QED anomalous term in the PQ effective Lagrangian (\ref{lanomaly}).

The effective axion and two photons coupling 
$g^s_{a\gamma}$ is the basic parameter of the present sterile QCD axion model. 
It relates to the effective coupling ${\mathcal G}^\gamma_R$ (\ref{rhc0}) 
between photon and right-handed (sterile) neutrinos
,
\begin{eqnarray}
g^s_{a\gamma}=  ({\mathcal G}^\gamma_R)^2 \approx ({\mathcal G}_R)^2< 10^{-8}.
\label{eapc}
\end{eqnarray}
This absolutely upper limit $10^{-8}$ on $g^s_{a\gamma}$ comes 
from the situation (a) ${\mathcal G}^\gamma_R\sim{\mathcal G}_R \sim 10^{-4}$ (\ref{a}) and the 
$\alpha$ precision measurement (\ref{ggamma}). Other two possible situations: (b) $g^s_{a\gamma}< 10^{-12}$ for 
${\mathcal G}_R \sim 10^{-6}$ (\ref{b}) and (c) $g^s_{a\gamma}< 10^{-14}$ for 
${\mathcal G}_R \sim 10^{-7}$ (\ref{c}), inferred from the Xenon1T experiment \cite{Aprile2020a} and reference \cite{Shakeri2020}, 
see also Table \ref{limits}.   
Moreover, we should point out that it is possible for
${\mathcal G}^\gamma_R < {\mathcal G}_R$. In fact, the effective coupling 
${\mathcal G}^\gamma_R$ in the effective right-handed interacting 
Lagrangian (\ref{rhc0}) has to be determined or constrained by more experiments and observations, such as those for probing 
axion-like particles studied in the present article.

\begin{figure}
\centering
\includegraphics[width=5.5in]{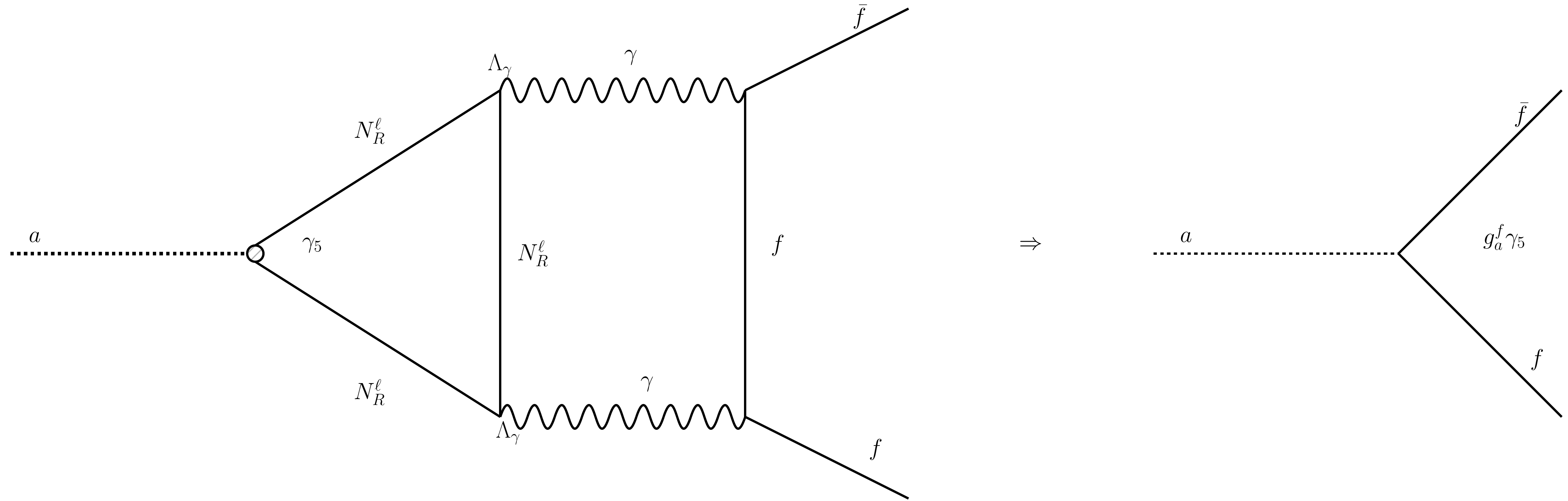}
    \caption{Left: the possible Feynman diagram represents leading-order effective coupling of an axion $a$ and two SM fermions $\bar f f$. 
    This diagram is the same as the boxed diagram of Figure 1 in Ref.~\cite{Kim1979} of the KSVZ model, 
    except two replacements: (i) the super heavy quark triangle loop by the 
massive sterile neutrino $N^\ell_R$ triangle loop, (ii) two gluon lines by two photon lines. Right: the induced 
1PI axial Yukawa coupling $g_a^f$ (\ref{ayukawa}) or (\ref{gfa}) of an axion $a$ and two SM fermions $\bar f f$.}
    \label{yukawaf}
\end{figure}


The anomalous 1PI vertex (\ref{panomaly}) induces an effective axial Yukawa coupling between the axion $a$ and SM fermions $f=q,\ell$, see Fig.~\ref{yukawaf}, which can be estimated as,
\begin{eqnarray}
i\alpha^2 g^s_{a\gamma} \left(\frac{m_{_f}}{f_a}\right)\ln\left(\frac{m^M_3}{m_{_f}}\right)(\bar f\gamma_5 f)a,
\label{ayukawa} 
\end{eqnarray}
where $m^M_3$ is the heaviest sterile neutrino Majorana mass and 
$m_{_f}$ is SM fermion masses. 
Such axion-Yukawa coupling (\ref{ayukawa}) is analogous to the result (6) 
in Ref.~\cite{Kim1979} of the KSVZ model, where a heavy electroweak singlet quark is introduced. As a result,
\comment{
Using the interaction (\ref{bhlbv}), 
we calculate scalar boson $\chi$ and 
pseudo scalar 
boson $a$ coupling to SM quarks and leptons, as represented in Fig.~\ref{coupling}. 
The composite pseudo scalar boson 
$a$ or scalar boson $\chi$ (the dashed line) couples to a sterile neutrino bubble, which couples 
to SM fermion via four-fermion interaction (\ref{bhlbv}). Using the
tadpole diagram of Fig.~\ref{tadpole}, Viz the gap equation 
$(G/2)\Pi_{s,p}(\mu^2_{s,p})=1$, we obtain in Fig.~\ref{coupling} an effective coupling 
for the scalar channel, and $ig^f_a\bar f\gamma_5 fa$ for pseudo scalar channel and $ig^f_s\bar f f \chi$.
}  
the corresponding effective Lagrangian is 
\begin{eqnarray}
{\mathcal L}_{\rm eff} &\supset& i\sum_q g^q_a(\bar q\gamma_5 q)
a+ i\sum_\ell g^\ell_a(\bar \ell\gamma_5 \ell) a
\label{layukawa}
\end{eqnarray}
where the sum is over all SM quarks or charged leptons, and $(\bar q\gamma_5 q)\equiv (\bar q^c\gamma_5 q_c)$ is a color singlet, the same below. 
The axion Yukawa couplings $g^{q,\ell}_a$ to SM quarks and 
charged leptons $f=q,\ell$ are given by 
\begin{eqnarray}
 g^f_a&=&\alpha^2 g^s_{a\gamma} \left(\frac{m_{_f}}{f_a}\right)\ln\left(\frac{m^M_3}{m_{_f}}\right) \ll 1,
\label{gfa}
\end{eqnarray}
and $g^f_a\ll 1$ is due to $g^s_{a\gamma}\ll 1$ (\ref{eapc}). 
This axial Yukawa coupling 
(\ref{gfa}) is proportional to SM fermion mass $m_f$, and 
$\ln (m^M_3/m_{_f})$ is a slowly varying function of $m_f$. Thus the axion decay to two SM fermions should be dominated by heavy fermion channels. Considering the top-quark channel ($m_f=m_t$) and assuming 
the heaviest sterile neutrino mass $m^M_3=v$, we have the absolutely upper limit of the largest axion-fermion Yukawa coupling    
\begin{eqnarray}
 g^t_a&=&\alpha^2 g^s_{a\gamma} \left(\frac{m_{_t}}{f_a}\right)\ln\left(\frac{m^M_3}{m_{_t}}\right) < 1.88\times 10^{-13}.
\label{gfat}
\end{eqnarray}
In Table \ref{limits}, we tabulate the upper limits of $g^t_a$ 
for all situations (a), (b) and (c) inferred from the Xenon1T experiment \cite{Aprile2020a} and reference \cite{Shakeri2020}.  

\comment{Based on the axial coupling (\ref{layukawa}) and triangle diagram (left) of Fig.~\ref{trianglea}, we rewrite the axial couplings (\ref{gfa}) as
\begin{eqnarray}
g^{q,\ell}_a=(m_{q,\ell}/f_a)\xi^{q,\ell}, 
\quad \xi^{q,\ell} 
\equiv \alpha^2 g^s_{a\gamma}\ln\left(\frac{m^M_3}{m_{q,\ell}}\right) \ll 1.
\label{gfa1}
\end{eqnarray}
}

\subsubsection{Axion mass and coupling to QCD anomaly
}\label{axionQCD0}

\begin{figure}[t]
\centering
\includegraphics[width=2.5in]{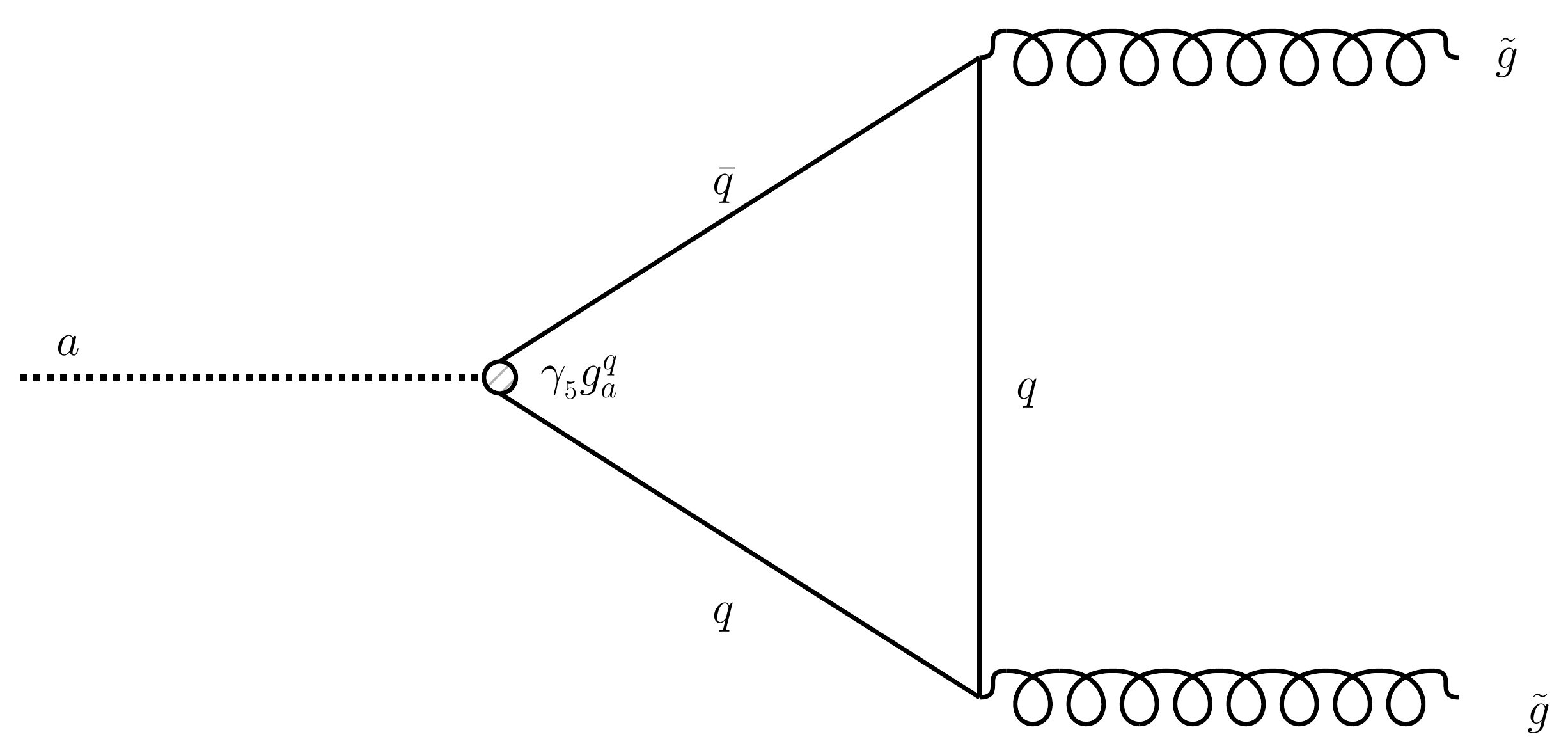}\hskip1.5cm
\includegraphics[width=2.7in]{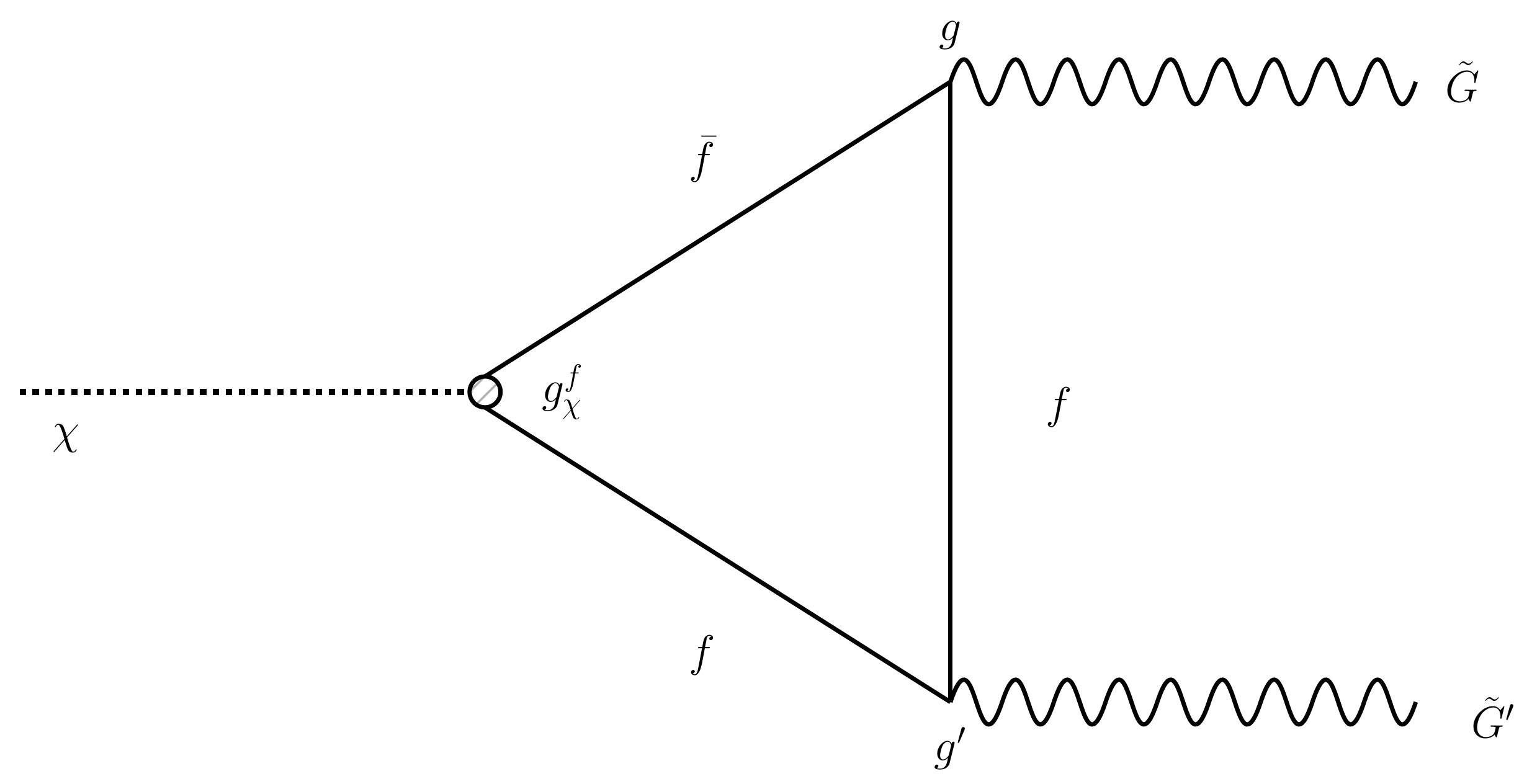}
\caption{Left: The axial current $J^\mu_{PQ}$ anomaly (\ref{canomaly}) or 
		(\ref{lanomaly}) is obtained by calculating this triangle quark-loop diagram in usual manner. The axial coupling $g^f_a=g^q_a$ (\ref{gfa}) and two gluons are represented by $\tilde g$ and $\tilde g$. This shows that 
$a$ is the PQ QCD axion. Right: The $\chi$boson decay to two SM gauge bosons $\tilde G$ and $\tilde G'$ via this SM fermion loop triangle 
diagram.  The coupling $g^f_\chi=m_f/f_a$ is the scalar coupling 
of $\chi$boson and SM fermions. The SM gauge couplings $g$ and $g'$ 
relate to gauge bosons $\tilde G\tilde G'= \gamma\gamma, \gamma Z^0, Z^0Z^0, 
W^+W^-$ and two gluons.}
\label{trianglea}
\end{figure}

Now we see how to achieve the QCD anomalous term in the effective 
PQ Lagrangian (\ref{lanomaly}). 
Using the axion and quarks axial coupling (\ref{layukawa},\ref{gfa}) and employing standard approach to calculate axial anomaly, i.e., the anomalous triangle diagram (left) of Fig.~\ref{trianglea}, we obtain the anomalous QCD term 
\begin{eqnarray}
{\mathcal L}_{\rm eff} &\supset& \xi_s \frac{a}{f_a}\frac{g_s^2}{32\pi^2}F^a_{\mu\nu}\tilde F_a^{\mu\nu},
\label{qcda} 
\end{eqnarray}
in the PQ effective Lagrangian (\ref{lanomaly}).   
As a result, we determine the model-dependent QCD axion coefficient $\xi$ for the sterile QCD axion model
\begin{eqnarray}
\xi_s = \alpha^2 g^s_{a\gamma} \sum_q \ln\left(\frac{m^M_3}{m_q}\right). 
\label{pqxi} 
\end{eqnarray}
This result $\xi_s\propto g^s_{a\gamma}\ll 1$ is different 
from the parameter-free PQ axion model $\xi\sim {\mathcal O}(1)$ (\ref{pqxi0}), though the PQ symmetry breaking scale is the same $f_a=v$ 
at electroweak one. 
This result is also different from invisible axion models 
of KSVZ and DFSZ types with PQ symmetry breaking scale $f_a^{\rm inv}\gg v$. The $\xi_s$ smallness is due to the tiny coupling of photon and right-handed neutrino ${\mathcal G}_R^\gamma$ (\ref{rhc0}), see also (\ref{eapc}), in contrast with the suppression by a newly introduced high-energy scale $f_a^{\rm inv}\gg v$ in invisible axion models. The common feature of sterile QCD axion and invisible QCD axion models is one free parameter model. The former is the photon and sterile neutrino coupling ${\mathcal G}_R^\gamma$, and the latter is the PQ symmetry breaking scale 
$f_a^{\rm inv}$.    
 
We cannot determine the factor $\sum_q\ln (m^M_3/m_q)$, which is expected to be $\sim {\mathcal O}(1)$ or at most $\sim {\mathcal O}(10)$. The reason is that the heaviest sterile neutrino $N_R^\tau$ ($m^M_3\sim 10^2$ GeV) gives main contribution to the triangle diagrams of Figs.~\ref{nnupf} and \ref{yukawaf}, thus $\ln (m^M_3/m_{u,d})<10$. Then,
from the $g^s_{a\gamma}$ upper limit (\ref{eapc}), 
the absolute upper limit on $\xi_s$ yields,  
\begin{eqnarray}
\xi_s < 10^{-12}
\label{pqxi1} 
\end{eqnarray}
at least for (a) $g^s_{a\gamma}< 10^{-8}$ in the situation (\ref{a}). 
Whereas (b) $\xi_s < 10^{-16}$ and $g^s_{a\gamma}< 10^{-12}$ 
in the situation (\ref{b}); $\xi_s < 10^{-18}$ and $g^s_{a\gamma}< 10^{-14}$ in the situation (\ref{c}), see Table \ref{limits}. 

Once the dimensionless coefficient $\xi_s$ is determined, following the
PQ QCD axion approach in Refs.~\cite{Bardeen1978,BARDEEN1987401} 
and \cite{Kim1979}, 
we can approximately calculate 
the axion mass $m_a$, 
\begin{eqnarray}
m_a\approx \xi_s m_\pi \frac{f_\pi}{f_a}\frac{\sqrt{m_um_d}}{m_u+m_d},
\label{amasss} 
\end{eqnarray}
for the sterile QCD axion model. This relation is similar to the 
relation (\ref{amass}) for the PQ axion model, changing 
the dimensionless coefficient from $\xi$ (\ref{pqxi0}) 
to $\xi_s$ (\ref{pqxi}). It is also similar to the one obtained in 
Ref.~\cite{Kim1979}, changing the coupling from the QCD 
$\alpha_s$ to the QED $\alpha$.

Corresponding to the axial coupling (\ref{layukawa}), the axial current of sterile neutrino $U^{\rm PQ}_{\rm lepton}(1)$ symmetry is 
$J^\mu_{PQ}=\delta {\mathcal L}_{\rm eff}/\delta (\partial_\mu a)$,
\begin{eqnarray}
J^\mu_{PQ} = -f_a\partial^\mu a + \sum_q g^q_a(\bar q\gamma_5 \gamma^\mu q)
+\sum_\ell g^\ell_a(\bar \ell\gamma_5 \gamma^\mu\ell).
\label{pqcurrent} 
\end{eqnarray}
This axial current has the correct PQ chiral anomaly (\ref{canomaly}).

It should also be mentioned that the axial coupling (\ref{layukawa}) 
of axion and SM leptons gives the next order correction 
to the axion-photon coupling (\ref{panomaly})
\begin{eqnarray}
g^s_{a\gamma}\rightarrow g^s_{a\gamma}\Big[1+\alpha^2\sum_\ell \ln (m^M_3/m_{_\ell})\Big].
\label{pqga} 
\end{eqnarray}
This is obtained by calculating the triangle diagram (Right) 
of Fig.~\ref{nnupf} in the standard way. 

\subsubsection{Sterile QCD axion at electroweak scale 
}\label{axionQCD}

This sterile QCD axion model provides the PQ solution 
to the strong CP problem of QCD in the usual manner described 
in Sec.~\ref{QCDaxion}. Observe that the QCD CP-violation 
parameter $\bar\theta \ll 1$ and the sterile QCD-axion coefficient 
$\xi_s\ll 1$ (\ref{pqxi}). This implies the possibility 
$\bar\theta\sim \xi_s$ that the sterile-neutrino composed 
QCD axion condensation 
$|\langle a\rangle |=\bar \theta f_a/\xi_s \sim f_a$ 
is the same order of magnitude of electroweak scale $f_a=v$. 

Due to the smallness of sterile neutrino coupling (\ref{rhc0}) 
or (\ref{gR0}), the axion-photon 
coupling $g^s_{a\gamma}$ (\ref{eapc}), axion-fermion Yukawa 
coupling (\ref{gfa}), the sterile QCD-axion coefficient $\xi_s$ (\ref{pqxi}) 
and axion mass $m_a$ (\ref{amass}) are very small, even for 
the PQ symmetry $U^{\rm PQ}_{\rm lepton}(1)$ breaking scale $f_a$ 
is the same as the electroweak scale $v$. 
This result is drastically different from the PQ original model, 
whose the axion-photon coupling $\xi\sim {\mathcal O}(1)$ (\ref{pqxi0}).
Moreover, the sterile QCD-axion scenario has a striking difference 
from the KSVZ and DFSZ models, where the couplings of axion and other 
SM particles are suppressed instead by a high-energy scale $f_a\sim 10^{11}$ GeV. Nevertheless, the sterile-neutrino composed axion discussed here is just a QCD axion of PQ type, 
rather than an extra axion-like particle (ALP).  

The physical axion acquires its mass $m_a$ at the minimum of axion potential, and can also receive the contribution from the explicit breaking of the PQ chiral symmetry $U^{\rm PQ}_{\rm lepton}(1)$. 
Namely, generated by explicit chiral symmetry breaking (\ref{mam1}),
sterile neutrino mass terms $m^M_{1,2}$ (\ref{lmass}) contribute to the axion mass $m_a$. This is analogous to the pion mass $m_\pi$ receives contributions from the $u$ and $d$ quark masses.
In this case, anomalous current conservation (\ref{canomaly}) is modified to
\begin{eqnarray}
\partial_\mu J^\mu_{PQ} =f_a m_a^2a+ \xi_s\frac{ g^2_s}{32\pi^2
}F_{\mu\nu}^a\tilde F^{\mu\nu}_a + g^s_{a\gamma}\frac{e^2}{32\pi^2}F_{\mu\nu}\tilde F^{\mu\nu},
\label{canomaly1} 
\end{eqnarray} 
where the term $f_a m_a^2a$ is added. This is similar to the partial conservation of axial current (PCAC) for the $\pi$ meson Physics.

\subsection{Sterile QCD axion candidate for superlight dark matter particle}
\label{sga}

In the present sterile QCD axion model, the induced 1PI effective operators of axion couplings to SM particles (\ref{layukawa}) and (\ref{qcda}) violate the lepton number conservation, since such a sterile axion $a$ carries 
two units of lepton (PQ) numbers. Any lepton-number violating process 
due to these 1PI effective operators is highly suppressed. The reason is that 
the sterile QCD axion coefficient $\xi_s$ (\ref{pqxi}), the axion-photon 
coupling $g^s_{a\gamma}$ (\ref{eapc}) and axion-fermion couplings 
$g^{q,\ell}_a$ (\ref{gfa}) 
are very small. This implies that the interactions between the 
sterile sector 
$(N^\ell_R,a,\chi)$ and the SM particles are very tiny.

Now we are in the position of studying the parameter space of axion coupling $(g^{s,{\rm exp}}_{a\gamma})$ and axion mass 
$(m_a)$ for the sterile QCD axion model. 
From the absolutely upper limit $g^s_{a\gamma}< 10^{-8}$ (\ref{eapc}) for the situation (a) ${\mathcal G}_R \sim 10^{-4}$ (\ref{a}), we use 
Eqs.~(\ref{pqxi1}) and (\ref{amasss}) to obtain the absolutely upper limits of experimentally related axion mass $m_a$ and  
axion-photon coupling $g^{s,{\rm exp}}_{a\gamma}$
\begin{eqnarray}
m_a&\approx& \xi_s m_\pi \frac{f_\pi}{f_a}\frac{\sqrt{m_um_d}}{m_u+m_d} < 10^{-8} {\rm eV},\nonumber\\
g^{s,{\rm exp}}_{a\gamma} &\equiv& g^s_{a\gamma}\frac{\alpha}{2\pi f_a} < 10^{-13} {\rm GeV}^{-1},
\label{apexp} 
\end{eqnarray}
where $f_a\approx 246$GeV and $\alpha =1/137$.
Other two possible situations (b) and (c) can be found in Table \ref{limits}.
These upper bounds (\ref{gfat}) and (\ref{apexp}) 
are below the limits reached by current laboratory experiments and astrophysical observations. For example 
$g_{a\gamma}<1.4\times 10^{-10}{\rm GeV}^{-1}$ and $3.1\times 10^{-10}< m_a <8.3\times 10^{-9}$ eV from the ABRACADABRA experiments \cite{Ouellet2018}, and at $m_a\sim 20$ peV reaching $4.0\times 10^{-11}{\rm GeV}^{-1}$ 
\cite{Gramolin2020}. These results are competitive with the most stringent astrophysical constraints in these mass ranges. The future experiments 
\cite{Obata2018,Michimura2020,Nagano2019} are possible to probe the sterile QCD axion relation (\ref{amgag}) in the coupling-mass range 
bound by Eqs.~(\ref{gfat}) and (\ref{apexp}).
 
\begin{table*}
\begin{tabular}{cccccc}
$~~~~{\mathcal G}_R$ & $~g^s_{a\gamma}$ &$\xi_s$ & $~~~~~g_a^t$ &$~~~g^{s,{\rm exp}}_{a\gamma} ~[{\rm GeV}^{-1}]$ & $~~~m_a ~[{\rm eV}]$\cr
\hline
(a)~ $\sim 10^{-4} $& $~ < 10^{-8}$ & $~< 10^{-12}$ & $~~~~~< 10^{-13}$ & $~~< 10^{-13}$ &$~~< 10^{-8}$\cr
(b)~~$\sim 10^{-6} $ & $ ~~ < 10^{-12}$ & $~< 10^{-16}$& $~~~~~ < 10^{-17}$ &$~~< 10^{-17}$& $~~< 10^{-14}$\cr
(c)~~$\sim 10^{-7} $ & $ ~~ < 10^{-14}$ &$~< 10^{-18}$  & $~~~~~ < 10^{-19}$ &$~~< 10^{-19}$ &$~~< 10^{-16}$\\
\hline
\end{tabular}
\caption{We tabulate the constrains of parameters $g^s_{a\gamma}$ and 
$\xi_s$, and the upper limits of axion top quark coupling $g_a^t$, 
axion-photon coupling $g^{s,{\rm exp}}_{a\gamma}$ and axion mass $m_a$ in three situations (a), (b)and (c), see Eqs.~(\ref{a}), (\ref{b}) 
and (\ref{c}) inferred from the Xenon1T experiment \cite{Aprile2020a} and 
reference \cite{Shakeri2020}.}\label{limits}
\end{table*}

In addition, using the relations (\ref{pqxi}) and (\ref{amasss}), we obtain
the relationship between axion mass $m_a$ and axion-photon coupling $ g^{s,{\rm exp}}_{a\gamma}$ for the sterile QCD axion model
\begin{eqnarray}
m_a ({\rm eV}) &=& 2.69\times 10^{5} \sum_q \ln\left(\frac{m^M_3}{m_{q}}\right) g^{s,{\rm exp}}_{a\gamma} ({\rm GeV}^{-1}), 
\label{magag} 
\end{eqnarray}
where the $g^{s,{\rm exp}}_{a\gamma}$ is defined by Eq.~(\ref{apexp}). 
The difference between the relation (\ref{magag}) and the 
PQ relation (\ref{amgag1}) comes from the difference between $\xi=\xi(g_{a\gamma})$ (\ref{amass}) and $\xi_s=\xi_s(g^s_{a\gamma})$ (\ref{pqxi}).
In the conventional parameter space 
($g^{s,{\rm exp}}_{a\gamma}, m_a$) of axion coupling and mass, 
we plot in Fig.~\ref{Bounds} the sterile QCD axion relation (\ref{magag}) 
for $\sum_q\ln (m^M_3/m_{q})\in [1-10]$. To compare and contrast, we also show in Fig.~\ref{Bounds} (left) the results of other QCD axion models, available data (solid lines) that exclude some parts of the parameter space, as well as proposed sensitivities (dashed lines) that will be reached by new or upgraded experimental and observational measurements.

As a result, we find that the existing data have not 
excluded the sterile QCD axion model for its absolutely 
upper limits on the axion coupling $g^{s,{\rm exp}}_{a\gamma}$ and mass 
$m_a$ (\ref{apexp}), see also Table \ref{limits}. 
Moreover, we point out that the sterile QCD axion model can 
be probably probed by the proposed sensitivity of upgraded ABRACADABRA experiments \cite{Ouellet2018} in the range of axion coupling $g^{s,{\rm exp}}_{a\gamma}\sim (10^{-14}-10^{-16}) ~
{\rm GeV}^{-1}$ and mass $m_a\sim (10^{-8} - 10^{-11}) ~{\rm eV}$. To illustrate this possibility, we show more detailed plots in 
Fig.~\ref{Bounds} (right). 

\comment{Also we obtain the upper limits of sterile QCD axion coefficient $\xi$ (\ref{pqxi}) and 
axion mass (\ref{amass})
\begin{eqnarray}
\xi < 10^{-11},\quad m_a< 10^{-6} {\rm eV},
\label{amass1} 
\end{eqnarray}
as well as axion axial Yukawa couplings to SM fermions (\ref{gfa}) 
\begin{eqnarray}
g^f_a 
< 10^{-10}
\label{gfa1}
\end{eqnarray}
for $m_{_f}\lesssim m^M_3$.}
  

These properties (\ref{gfat}) and (\ref{apexp}) indicate that 
the sterile axion $a$ couplings and decay rates to SM particles 
are so small. Therefore such axion has a lifetime longer than the Universe lifetime. Its role in normal astrophysical processes should be negligible. It participates in gravitating processes. This implies that the sterile axion could be a candidate for superlight dark matter particles. The extremely tiny axion mass gravitationally accounts for the formation, evolution and structure of the Universe at very large scales. However, to verify these situations, further studies are required. 

\section{
Sterile Higgs-like boson and 
massive dark matter particle 
}\label{Xboson}

We turn to another composite boson, massive $\chi$boson of $m_\chi\sim 10^2$ GeV (\ref{mchi}), as a consequence of broken 
$U^{\rm PQ}_{\rm lepton}(1)$ symmetry of sterile neutrinos. 
This massive scalar 
$\chi$boson is the counterpart of the Higgs boson in the electroweak symmetry breaking. It is absent in the usual QCD axion models. 
We study its effective couplings to
SM fermions and gauge bosons, so as to estimate its decay rate to SM particles and interacting cross-section with nucleons, 
in comparison with currently ongoing experiments detecting massive dark matter particles at the mass range $\sim 10^2$ GeV. 
Because of very tiny decay rates and interacting cross-sections, the 
$\chi$boson has a lifetime longer than the Universe, 
could be a candidate for massive 
dark matter particles. We examine the $\chi$boson 
decay rates and interacting cross-sections to SM particles and nucleons, in comparison with currently ongoing experiments detecting massive dark matter particles in the mass range of $\sim 10^2$ GeV. The knowledge of $\chi$boson coupling strength to SM particles is an essential prerequisite for understanding how the $\chi$boson couples to and decouples from the thermal state of SM particles in the early Universe. Thereby, we can further figure out which kind of dark matter particle relics, WIMP or else that the $\chi$boson can be a candidate for. This subject however will be discussed in a separate article. 

\subsection{Massive 
scalar boson couplings to SM particles}

Replacing the axion $a$ by the $\chi$boson and Dirac matrix $\gamma_5$ 
by unity $1$ in Figs.~\ref{nnupf} and \ref{yukawaf}, the similar discussions and calculations lead to the estimated 
coupling between the scalar $\chi$boson and two photons. The results
give rise to the 1PI contribution to the effective Lagrangian,  
\begin{eqnarray}
{\mathcal L}_{\rm eff} &\supset& g^s_{\chi\gamma}\frac{\chi}{f_a}\frac{e^2}{32\pi^2}F_{\mu\nu} F^{\mu\nu}.
\label{xanomaly} 
\end{eqnarray} 
This is analogous to the Higgs boson coupling to two photons via a triangle SM fermion loop, see for example Ref.~\cite{Shifman:1979eb}. 
We assume the effective coupling of $\chi$boson and two photons is not larger than $g^s_{a\gamma}$,
\begin{eqnarray}
g^s_{\chi\gamma}&\lesssim & g^s_{a\gamma} < 10^{-8},
\label{eapcx}
\end{eqnarray}
whose value in the order of magnitude is probably close to the value of the axion-photon coupling $g^s_{a\gamma}$ (\ref{eapc}). The experimentally related coupling is defined as,
\begin{eqnarray}
g^{s,{\rm exp}}_{\chi\gamma} &\equiv& g^s_{\chi\gamma}\frac{\alpha}{2\pi f_a} < 10^{-13} {\rm GeV}^{-1},
\label{xpexp}  
\end{eqnarray} 
where the same decay constant $f_a=v$ is at the electroweak scale.

The scalar Yukawa coupling between the $\chi$boson and SM fermions 
$f=q,\ell$ yields   
\begin{eqnarray}
i\alpha^2 g^s_{\chi\gamma} \left(\frac{m_{_f}}{f_a}\right)\ln\left(\frac{m^M_3}{m_{_f}}\right)(\bar ff)\chi.
\label{xyukawa} 
\end{eqnarray}
The corresponding effective Lagrangian is 
\begin{eqnarray}
{\mathcal L}_{\rm eff} &\supset& i\sum_q g^q_\chi(\bar q q)
\chi+i\sum_\ell g^\ell_\chi(\bar \ell \ell) \chi, \label{xlyukawa}
\end{eqnarray}
where the sum is over all SM quarks or charged leptons. The $\chi$boson 
scalar Yukawa couplings $g^{q,\ell}_\chi$ to SM fermions are given by 
\begin{eqnarray}
 g^f_\chi&=&\alpha^2 g^s_{\chi\gamma} \left(\frac{m_{_f}}{f_a}\right)\ln\left(\frac{m^M_3}{m_{_f}}\right)< 10^{-13}.
\label{gfx}
\end{eqnarray}
The absolutely upper limit of the coupling $g^f_\chi< 10^{-13}$ come from the estimates (\ref{eapcx}) and (\ref{xpexp}), as well as discussions below Eq.~(\ref{pqxi}), analogously to the axion case. Actually, 
these are  counterparts of the axion axial 
couplings (\ref{ayukawa}), (\ref{gfa}) and 
effective Lagrangian (\ref{layukawa}).

As a consequence of these non-vanishing gauge couplings, 
the massive $\chi$boson can decay to two photons and 
an SM fermion pair $\bar ff$. 
In addition to photon-pair $\gamma\gamma$,
the $\chi$boson can decay to SM gauge boson pair 
$\tilde G'\tilde G= \gamma Z^0$, $Z^0Z^0$, $W^+W^-$ 
and two gluons $\tilde g\tilde g$ via the triangle SM fermion-loop diagram, as illustrated in the right diagram of Fig.~\ref{trianglea}. These decays are only possible when the $\chi$boson mass $m_\chi$ is larger than the kinematic thresholds of decay channels. The effective contact interacting Lagrangians for the decay channel $\chi\rightarrow \tilde G'\tilde G$ is the same as the one (\ref{xanomaly}) for the photon pair channel $\chi\rightarrow \gamma\gamma$. However, the electric coupling $e^2$ in Eq.~(\ref{xanomaly}) should be replaced by gauge 
coupling $gg'$ associating to two gauge bosons 
$\tilde G\tilde G'$ in final states \footnote{The decay rates of various channels can be obtained, as an analogy, see Eqs.~(80-83) for the heavy pion-like composite boson decay at TeV scale in Ref.~\cite{Xue:2016txt}}. These decay and interacting channels with the final state of gamma rays play a pronounced role among the various possible messengers
from massive 
dark matter particles \cite{Bergstrom1997a,Ullio2002a,
Bringmann2012a}.

\subsection{
Possible ways to probe sterile Higgs-like
scalar boson}

The induced 1PI effective operators of $\chi$boson couplings to SM particles 
(\ref{xanomaly}) and (\ref{xlyukawa}) violate the 
lepton number conservation, since such a $\chi$boson carries two units 
of lepton (PQ) numbers. Lepton-number violating processes 
due to these 1PI effective operators are highly suppressed, 
as the $\chi$boson-photon coupling $g^s_{\chi\gamma}$ (\ref{xanomaly}) and 
axion-fermion couplings $g^f_\chi$ (\ref{gfx}) 
are very small. 

These effective 1PI operators of $\chi$boson 
have the same form as the operators of 
Higgs boson couplings to SM fermions and gauge bosons. 
However, compared with Higgs and SM fermions couplings $g^f_{_H}=(m_f/v)$, the $\chi$boson and 
SM fermions couplings $g^f_{\chi}$ (\ref{gfx}) is much smaller by a factor
\begin{eqnarray}
(g^f_\chi/g^f_{_H})&=&\alpha^2 g^s_{\chi\gamma} \ln(m^M_3/m_{_f}) < 10^{-13}.
\label{gfxs}
\end{eqnarray}
Therefore, the decay rates of $\chi$boson to 
SM particles are at least $(g^f_\chi/g^f_{_H})^2$ 
smaller than the counterparts of Higgs boson decay channels. For instance, 
the $\chi$boson decay rates are given by
\begin{eqnarray}
 \Gamma(\chi\rightarrow \bar f f)&=&(g^f_\chi/g^f_{_H})^2(m_\chi/m_{_H})\Gamma(H\rightarrow \bar f f),
\label{gcdff}\\
\Gamma(\chi\rightarrow \tilde G'\tilde G)&=&(g^f_\chi/g^f_{_H})^2(m_\chi/m_{_H})^3\Gamma(H\rightarrow \tilde G'\tilde G),
\label{gcdgg}
\end{eqnarray}
where $\Gamma(H\rightarrow \bar f f)\propto G_F m_f^2m_{_H}$ 
\cite{Braaten1980,DREES1990455}
and $\Gamma(H\rightarrow \tilde G'\tilde G)\propto \alpha^2G_F m_{_H}^3$ are 
the rates of the Higgs decay via SM fermion and boson channels \cite{Shifman:1979eb,ELLIS1976292,Marciano_2012}. 
\comment{11.  S.G. Gorishny, A.L. Kataev, S.A. Larin and L.R. Surguladze, Mod. Phys.Lett.A5(1990) 2703;  Phys. Rev.D43(1991) 1633;  A.L.  Kataev andV.T.  Kim,  Mod.  Phys.  Lett.A9(1994)  1309;  L.R.  Surguladze,  Phys.Lett.341(1994) 61; K.G. Chetyrkin, Phys. Lett.B390(1997) 309.}

As consequences, it gives a further constraint on the coupling $g^s_{\chi\gamma}$ value for the $\chi$boson lifetime $\tau_\chi = \Gamma_\chi^{-1}$ 
being longer than $\sim 4.4\times 10^{17}$ seconds of Universe age 
\cite{Audren2014}. This implies that the massive $\chi$boson could be a candidate for 
dark matter particles of masses $\sim {\mathcal O}(10^2)$ GeV, provided that its interactions with SM particles are much weaker than electroweak
interactions.
However, we need to also consider the possibilities of $\chi$boson direct decays to sterile neutrinos. Figure \ref{figys} shows 
$\bar g_{s}\sim {\mathcal O}(1)$ for the Yukawa coupling 
$\bar g_{s}\bar N^{3 c}_R N^{3}_R \chi$ (\ref{effsax}), but 
$\chi$boson cannot directly decay to two sterile neutrinos $\bar N^{3 c}_R$ and $N^{3}_R$ pair. Because the $\chi$boson mass $m_\chi$ is smaller than the pair mass $2m^M_3$, see Sec.~\ref{neutrinoM}.
\comment{If kinematically permitted, the massive $\chi$boson possibly decay to two light sterile neutrinos via the Yukawa coupling $\bar g^{(1,2)}_{s}\bar N^{(1,2) c}_R N^{(1,2)}_R \chi$ (\ref{effyuk}). The decay rate can be estimated as $\sim (\bar g^{(1,2)}_{s})^2 M_\chi$. Therefore the Yukawa couplings is constrained by $(\bar g^{(1,2)}_{s})^2< $  lepton number violation .... It implies ??? } 
\comment{
Therefore, we have to study its dynamics and state in the early Universe, 
to understand the relic abundance of the $\chi$boson that contributes to the total relic abundance of dark matter particles observed in the present time. In doing so, we might be able to see any relation between the $\chi$boson and WIMP dark matter particles.   
This is
a subject for future studies.
} 
    
Moreover, given the $\chi$boson mass $m_\chi\sim 10^2$ GeV and the coupling of $\chi$boson and SM fermions $g_\chi^q$ (\ref{gfx}), the cross section 
of $\chi$boson scattering off nucleon ($u,d$ quarks) can be estimated as 
$\sigma_{\chi n}\sim (g_\chi^q)^2/m_\chi^2< 10^{-58}{\rm cm}^2$. 
Similarly, the estimated cross section of $\chi$boson annihilation $\chi\chi\rightarrow \bar ff$ \cite{Steigman2012} is the same order of magnitude ${\mathcal O}[(g_\chi^q)^2/m_\chi^2]$. 
These cross-sections are so small that they are far below the limits reached by current LHC and underground dark matter experiments, for example, $\sigma_{\chi n}\sim 10^{-49}{\rm cm}^2$  at $10^2$ GeV \cite{Roszkowski2018}. These cross-sections are also below the neutrino floor 
of neutrino coherent scattering background \cite{OHare2020a}. 
Therefore, massive $\chi$boson is not expected to give important contributions to astrophysical processes. However, it should play important roles in self-gravitating processes of relevant length scale, e.g., 
dwarf galaxy formations. 

To end this section, for reader's convenience, we summarize the properties of the $\chi$boson in Table \ref{limitx}, analogously to  
Table \ref{limits} for the sterile QCD axion.  We note that the $\chi$boson mass $m_\chi\sim {\mathcal O}(10^2)$ GeV and coupling $g^{s,{\rm exp}}_{\chi\gamma} < 10^{-13}$ GeV$^{-1}$. However, it should be pointed out that there is not a relation between the $\chi$boson coupling $g^{s,{\rm exp}}_{\chi\gamma}$ and mass $m_\chi$, like the one for the sterile QCD axion relation (\ref{amgag}). 
It would be interesting to see the possibility to probe
the $\chi$boson based on its effective electromagnetic 
interaction (\ref{xanomaly}) and by using  special methods 
that are adopted or proposed to detect the 
axion \cite{Ouellet2018,Gramolin2020,Obata2018,Michimura2020,Nagano2019}.

\begin{table*}
\begin{tabular}{cccccc}
$~~~~{\mathcal G}_R$ & $~g^s_{\chi\gamma}$ &$g_\chi^t$ & $~~~~~\sigma_{\chi n}[{\rm cm}^2]$ &$~~~g^{s,{\rm exp}}_{\chi\gamma} ~[{\rm GeV}^{-1}]$ & $~~~m_\chi ~[{\rm GeV}]$\cr
\hline
(a)~ $\sim 10^{-4} $& $~ < 10^{-8}$ & $~<10^{-13}$ & $~~~~~< 10^{-58}$ & $~~< 10^{-13}$ &$~~ \sim 10^{2}$\cr
(b)~~$\sim 10^{-6} $ & $ ~~ < 10^{-12}$ &$~< 10^{-17}$& $~~~~~ < 10^{-66}$ &$~~< 10^{-17}$& $~~\sim 10^{2}$\cr
(c)~~$\sim 10^{-7} $ & $ ~~ < 10^{-14}$ &$~< 10^{-19}$  & $~~~~~ < 10^{-70}$ &$~~< 10^{-19}$ &$~~\sim 10^{2}$\\
\hline
\end{tabular}
\caption{We tabulate the constrains of parameters $g^s_{\chi\gamma}$ and the upper limits of $\chi$boson top quark coupling $g_\chi^t$, 
$\chi$boson-nucleon cross section $\sigma_{\chi n}$,
$\chi$boson-photon coupling $g^{s,{\rm exp}}_{\chi\gamma}$ and $\chi$boson mass $m_\chi$ in three situations (a), (b)and (c), see Eqs.~(\ref{a}), (\ref{b}) and (\ref{c}) inferred from the Xenon1T experiment \cite{Aprile2020a} and 
reference \cite{Shakeri2020}.}\label{limitx}
\end{table*}

\section
{Summary and remarks}\label{con}
\comment{As consequences, the $\chi $boson
cannot decay to SM particles, provided its coupling $g^s_{\chi\gamma}\ll 1$ and lifetime is much longer than
$\sim 4.4\times 10^{17}$ seconds of Universe age or
$\sim 200 {\mathrm{Gyr}} \approx 8.3\times 10^{18}$ seconds
\cite{Audren2014}. This implies that the massive $\chi $boson could be
a candidate for dark matter particles of masses
$\sim {\mathcal O}(10^{2})$ GeV. 
This implies that the massive $\chi $boson could be
a candidate for dark matter particles of masses
$\sim {\mathcal O}(10^{2})$ GeV.
}
The article presents a theoretical framework of the spontaneous PQ symmetry breaking of the operator (\ref{bhlxl}) of right-handed neutrinos 
$\nu^\ell_R$ and their induced interactions (\ref{rhc0}) 
with SM particles. We discuss three possible types of DM particle candidates: (i) the sterile neutrinos $N_R^{1,2,3}$ of masses ${\mathcal O}(10^2)$ keV, ${\mathcal O}(10^2)$ MeV and ${\mathcal O}(10^2)$ GeV; (ii)the superlight pseudoscalar axion $a$ of $m_a< 10^{-8}$ eV; (iii) the massive scalar $\chi$boson of $m_\chi \sim {\mathcal O}(10^2)$ GeV, and later ones are composed by formers. The constraints from $W$-boson decay width, the Xenon1T and precision fine-structure-constant $\alpha$ experiments give the upper limits of their indirectly induced interacting strengths to SM particles: (i) 
$\nu_R^\ell$ couplings to SM gauge bosons (\ref{rhc0}); (ii) $a$ axion couplings to SM gauge bosons (\ref{panomaly}) 
and fermions (\ref{layukawa}); (iii) $\chi$boson couplings to SM gauge bosons (\ref{xanomaly}) and fermions (\ref{xlyukawa}). These upper limits 
listed in Tables \ref{limits} and \ref{limitx} allow us to preliminarily estimate the lifetimes of these sterile particles, being larger than Universe age. This is the necessary but not sufficient condition for them to be possible DM particle candidates. In addition, these upper limits have to be consistent with the facts of their negligible contributions to all known astrophysics processes, current laboratory experiments and astrophysical observations for directly or indirectly detecting these DM particle candidates.  

In such a taxonomy of DM particle candidates, 
their different mass scales could qualitatively account for gravitational effects at different length scales. However, to give a quantitative description of these effects, one may need to know the self-interactions
of these DM particles, see for example Ref.~\cite{Arvanitaki2019}.  
It should be mentioned that at the high order of the large-$N$ expansion 
in Sec.~\ref{sterile}, one can obtain the axion and $\chi$boson self-interactions. Also, the high-order contributions of coupling ${\mathcal G}_R$ between DM and SM particles can induce the 1PI self-interacting of these dark matter particles. 

Moreover, whether or not the relic densities of these DM particle candidates can account for the gravitational effects observed in galactic 
and cosmological scales. How to understand their relic abundances that contribute to the total relic abundance of DM particles observed in the present time. To answer these questions, we have to use these DM particle candidates' mass spectra and their couplings to SM particles 
in Boltzmann rate equations to study their productions, thermodynamical states and evolutions in the early Universe, and when they decouple from thermal equilibrium or energy equilibration with SM particles. For example, we might be able to see any relation between the $\chi$boson and WIMP dark matter particles.   
\comment{Then we have the question of how the ``WIMP miracle'' happens. 
play the role of the WIMP dark matter particles.
}  
Moreover, we have to mention the possibility of the fourth type of DM particle candidates, being very massive and much heavier than 
the $\chi$boson mass.  
They are gravitationally produced \cite{Chung2001,Chung2019,Ema2018,Hashiba2019,Hashiba2019a,Li2019,Xue2019,Xue2020,Xue2020b,Xue2020a,Xue2021,Xue2022}, and could be candidates for the cold dark matter (CDM). 

All DM particle candidates contribute to the total relic 
abundance of dark matter $\Omega_{DM}h^{2}=0.120\pm 0.001$ observed 
today \cite{Aghanim2020,Allahverdi:2020bys,Boyarsky:2018tvu}. 
This demands that the contribution from each DM particle candidate should
be smaller than this amount. 
The natural question is then how much contribution from each DM particle candidate proposed in this article. 
\comment{
This certainly depends on how the DM particle candidates are produced, 
whether they were in thermal equilibrium or energy equilibration with other particles, and when they decouple in the early Universe. 
These should be consistently related to small couplings studied in this article.} 
These questions are subjects of future studies.

\section
{\bf Acknowledgment.}  
The author thanks S.~Shakeri and F.~Hajkarim for many discussions on sterile neutrinos and Xenon1T experiment results. 
The author also thanks the 
referee for the report that makes me improve the article.


\providecommand{\href}[2]{#2}\begingroup\raggedright\endgroup

\end{document}